\documentclass[aps,floats,prd,showpacs]{revtex4}
\usepackage{amsmath}
\usepackage{amsfonts,amsmath,mathrsfs}
\usepackage{epsfig}
\usepackage{latexsym}
\usepackage{bbm}
\usepackage{amssymb}
\usepackage{graphicx}
\usepackage{dcolumn}
\usepackage{bm}

\begin{document}

\title{Superconducting ground state of a doped Mott insulator }
\author{Zheng-Yu Weng}
\affiliation{\textit{Instutute for Advanced Study, Tsinghua University, Beijing 100084,
China}}

\begin{abstract}
A d-wave superconducting ground state for a doped Mott insulator is
obtained. It is distinguished from a Gutzwiller-projected BCS superconductor
by an explicit separation of Cooper pairing and resonating valence bond
(RVB) pairing. Such a state satisfies the precise sign structure of the t-J
model, just like that a BCS state satisfies the Fermi-Dirac statistics. This
new class of wavefunctions can be intrinsically characterized and
effectively manipulated by electron fractionalization with neutral spinons
and \textquotedblleft backflow\textquotedblright \ spinons forming a
two-component RVB structure. While the former spinon is bosonic, originated
from the superexchange correlation, the latter spinon is found to be
fermionic,\ accompanying the hopping of bosonic holons. The low-lying
emergent gauge fields associated with such a specific fractionalization are
of mutual Chern-Simons type. Corresponding to this superconducting ground
state, three types of elementary excitations are identified. Among them a
Bogoliubov nodal quasiparticle is conventional, while the other two are
neutral excitations of non-BCS type that play crucial roles in higher
energy/temperaure regimes. Their unique experimental implications for the
cuprates are briefly discussed.
\end{abstract}

\date{{\small \today}}
\maketitle

\section{Introduction}

An important issue in the study of high-$T_{c}$ cuprates concerns how
superconductivity can arise in a doped Mott insulator\cite{pwa_87}. The
d-wave pairing symmetry has been usually attributed to the reason that the
electrons avoid a strong local Coulomb repulsion. But in a doped Mott
insulator the on-site Coulomb repulsion is so strong that the Hilbert space
of the electrons is also drastically altered. It is thus no longer
sufficient just to focus on the relevant attractive interaction as in the
framework of the BCS theory. Rather a fundamental change in the underlying
electronic structure should be taken into account before one can
meaningfully address the issue of high-$T_{c}$ superconductivity. \

The simplest straightforward method of incorporating\ superconductivity with
the \textquotedblleft Mott physics\textquotedblright \ is to construct a
Gutzwiller-projected BCS superconductor. As first envisaged by Anderson\cite%
{pwa_87} in 1987, this class of state is always insulating at half-filling
as all the double-occupancy states get projected out and the original Cooper
pairs in the BCS wavefunction become neutralized, known as the spin-singlet
RVB pairs which are \textquotedblleft glued\textquotedblright \ by the
superexchange coupling (the state is to be referred to as the fermionic RVB
state below). Superconductivity arises only away from half-filling when the
RVB pairs start to move in the presence of, say, empty sites in the hole
doped case, which become partially charged Cooper pairs. Here the Cooper and
RVB pairings are no longer explicitly distinguishable.

The Gutzwiller-projected d-wave BCS state has been studied\cite%
{pwa_03,gros_07} intensively as a class of variational wavefunctions for the
doped Mott insulator. It is also the basis for developing the so-called
slave-boson approach\cite{LNW_06}, which is an electron fractionalization
description with the Gutzwiller projection replaced by emergent gauge
fluctuations around the spin-charge-separated saddle-points. It predicted%
\cite{bza_87,zhang_88} the presence of a high-temperature pseudogap phase
over a finite doping regime and a superconducting dome at lower
temperatures, which are both qualitatively consistent with the later
experimental measurements in the cuprates\cite{timusk_99,shen_03}.

However, the important long-range antiferromagnetic (AFM) correlations are
notably missing in the Gutzwiller-projected BCS state at half-filling. In
fact, it is a spin liquid with low-lying fermionic excitations\cite%
{pwa_87,bza_87,zhang_88}, which is in sharp contrast to a long-range AFM
order with bosonic spin-wave excitations governed by the two-dimensional
(2D) Heisenberg model. Of course one may argue here that once in the
presence of some finite concentration of doped holes, the long-range AFM
order or correlations will disappear anyway, and hence the
Gutzwiller-projected BCS/spin liquid state could become stabilized
eventually as a competitive ground state\cite{pwa_87,LNW_06}.

But it remains to be a real challenge to understand how the long-range AFM
order/correlations of an antiferromagnet/Mott insulator can be effectively
destroyed by the motion of the doped holes, and how the associated energy,
albeit only a small fraction in the total superexchange energy\cite{lda_88},
gets turned off in order to gain the kinetic energy of the doped holes. The
issue at the heart of a dope Mott insulator is the \emph{competition}
between the kinetic and superexchange energies -- if the AFM correlations
are weakened, like in the Gutzwiller-projected BCS state, it would be much
more favorable to the hopping of the doped holes; with the enhanced AFM
correlations at low doping, on the other hand, the kinetic energy of the
doped holes will get strongly suppressed. The novelty inherent from such
incompatibility and competition between the hopping and the superexchange
processes is thus expected\cite{weng_07} to be responsible for the
unconventional nature of the superconducting transition as well as a complex
pseudogap phenomenon over a wide temperature regime above $T_{c}$ in a doped
antiferromagnet/Mott insulator.

Therefore, to properly accommodate such novelty, which may provide a basic
understanding of the rich and marvelous pseudogap properties in the cuprates%
\cite{timusk_99,shen_03}, both the RVB and Cooper channels should remain
generally distinguished even in the superconducting regime, which implies
the necessity for one to go beyond the simple Gutzwiller-projected BCS state
approach to adequately address the interplay between the magnetism and
superconductivity.

Another important property that the wavefunction should obey is an altered
statistical sign rule: although the electrons always obey the Fermi-Dirac
statistics, which dictates that two electrons of the same spin cannot stay
at the same lattice site, new statistics\cite{zaanen_09} will emerge in a
doped Mott insulator where the no double occupancy constraint further
enforces that two electrons of \emph{opposite} spins cannot occupy the same
site. In fact, the t-J model at half-filling is totally \textquotedblleft
bosonized\textquotedblright \ in the restricted Hilbert space where the
usual fermion signs are completely diminished: e.g., the ground state only
possesses the trivial Marshall signs\cite{marshall_55,lda_88} which can be
easily gauged\ away. The nontrivial signs start to reemerge in the doped
case, induced by the hopping of the doped holes, which is precisely
described by the so-called phase string effect\cite{sheng_96,weng_97,WWZ_08}
in the t-J model. The corresponding sign structure is actually independent
of temperature, dimensionality, and is of statistical nature\cite%
{weng_97,WWZ_08}, which eventually recovers the full Fermi statistical signs
only at high doping in the dilute electron limit. Physically the phase
string effect also provides an accurate mathematical description of the
so-called \textquotedblleft unrenormalizable phase shift\textquotedblright \
first emphasized by Anderson early on\cite{pwa-book}. The latter is a total
phase shift added up from all the electrons in the ground state, in response
to adding/removing an electron into/from the system, due to the strong
on-site Coulomb repulsion. Consequently, the irreparable phase string
effect/unrenormalizable phase shift will make the Cooper pairing, associated
with the doped holes, \emph{intrinsically} distinguished from the neutral
spin RVB pairing caused by superexchange, again suggesting the necessity to
go beyond the Gutzwiller-projected BCS state description.

A superconducting ground state distinct from the Gutzwiller-projected BCS
state has been previously constructed by the present author and coauthors%
\cite{WZM_05} with incorporating the above-mentioned sign structure. Such a
ground state can naturally reduce to an insulating AFM state at
half-filling, which well accounts for the long-range AFM order as well as
the short-range spin-spin correlations with a highly accurate variational
superexchange energy, known as the bosonic RVB state\cite{lda_88,WZM_05}. In
contrast to the aforementioned fermionic RVB state, the doped holes are
quite unfavorable to hop in the bosonic RVB (neutral spin) background which
only involves the spin pairing between different sublattices. It was then
shown\cite{WZM_05} that the doped holes will force a fundamental change in
the RVB structure in order to gain the kinetic energy at finite doping,
characterized by emergent \textquotedblleft backflow
spinons\textquotedblright \ accompanying the hole hopping\cite{WZM_05}. It
is these \textquotedblleft backflow spinons\textquotedblright \ that will be
associated with the Cooper pairs instead of the original bosonic RVB
pairing. Consequently both the bosonic RVB and Cooper channels remain
explicitly separated in the superconducting state [cf. Eq. (84) in Ref. \cite%
{WZM_05}]. Such a wavefunction description has demonstrated a rich
complexity in the pseudogap regime as resulting from the competition between
the RVB and Cooper channels.

In this paper, we show that an important simplification in this approach can
be made by realizing that the aforementioned \textquotedblleft backflow
spinons\textquotedblright \ are actually fermionic upon a closer
reexamination of the sign structure. In the previous formulation, they are
described in the bosonic representation\cite{WZM_05}, which causes
unnecessary complications because the extra fermionic statistical signs are
mixed with the intrinsic phase string effect (cf. Sec. II A2). As the
result, we obtain a greatly simplified self-consistent description of both
the ground state and excitations for the doped Mott insulator.

The key results are summarized in Sec. II. The general form of the
superconducting ground state is presented in Sec. II A, which is
distinguished from the Gutzwiller-projected BCS superconductor by a novel
separation of Cooper and RVB pairings. It precisely satisfies the altered
statistical sign rule of the t-J model in the restricted Hilbert space,
which is of mutual semion type instead of the Fermi-Dirac one. In Sec. II B,
it is shown that such a new class of wavefunctions can be intrinsically
characterized by electron fractionalization, where neutral bosonic spinons
and \textquotedblleft backflow\textquotedblright \ fermionic spinons
together constitute a two-component RVB structure. The low-lying emergent
gauge fields associated with such a specific fractionalization are of mutual
Chern-Simons type, whose origin can be directly connected to the precise
sign structure of the t-J model. The fractionalization formalism also makes
the manipulation of the ground state significantly simplified as the
constituent subsystems are more conventional, governed by the effective
Hamiltonians presented in Sec. II B2. Finally, corresponding to this
superconducting ground state, three distinctive elementary excitations are
briefly discussed in Sec. II C. Among them a Bogoliubov nodal quasiparticle
is conventional, while the other two are non-BCS like neutral excitations
that play dominant roles in higher energy/temperature regimes, controlling
the superconducting phase transition and other exotic properties different
from a conventional d-wave BCS superconductor.

In Sec. III, a microscopic justification of the present approach is
presented in detail. It is based on a full bosonization formulation known as
the phase string representation\cite{weng_97} of the t-J model, in which the
whole nontrivial sign structure is explicitly captured by a topological
(mutual statistical) gauge structure. Then we show that such a formalism
under the no double occupancy constraint leads to the introduction of the
two-component spinons in order to adequately describe the microscopically
distinctive superexchange and hopping processes. Such a new exact
formulation provides a precise starting point that naturally results in an
electron-fractionalized description of the superconducting ground state at a
finite doping, and a highly accurate bosonic RVB description of the AFM
correlation in the zero doping limit. An effective theory of the elementary
excitations is also obtained within the same framework. Finally, the
conclusion and perspective are presented in Sec. IV.

\section{Key results}

We present this section to summarize the key equations/results of the
present work, which basically addresses the issue how the ground state of a
Heisenberg antiferromagnet/Mott insulator can be turned into a
superconducting ground state by doping.

\subsection{Ground state ansatz}

For comparison, let us start with the Gutzwiller-projected BCS state ansatz,
proposed\cite{pwa_87} for the t-J Hamiltonian on a 2D square lattice, given
by
\begin{equation}
|\Psi _{\mathrm{RVB}}\rangle =\hat{P}_{\mathrm{G}}|d\text{-}\mathrm{BCS}%
\rangle  \label{BCS}
\end{equation}%
where $|d$-$\mathrm{BCS}\rangle $ denotes an ordinary d-wave BCS state and $%
\hat{P}_{\mathrm{G}}$ is a Gutzwiller projection operator enforcing the
following no double occupancy constraint

\begin{equation}
\sum_{\sigma }c_{i\sigma }^{\dagger }c_{i\sigma }\leq 1.  \label{mottness}
\end{equation}%
Because of $\hat{P}_{\mathrm{G}}$, the Cooper pairing in $|d$-$\mathrm{BCS}%
\rangle $ reduces to the neutralized RVB pairing\cite{pwa_87} at
half-filling, whereas at finite doping the Cooper and RVB pairings are not
explicitly distinguished, as schematically illustrated in Fig. 1.
\begin{figure*}[bp]
\centering \includegraphics[width=12cm]{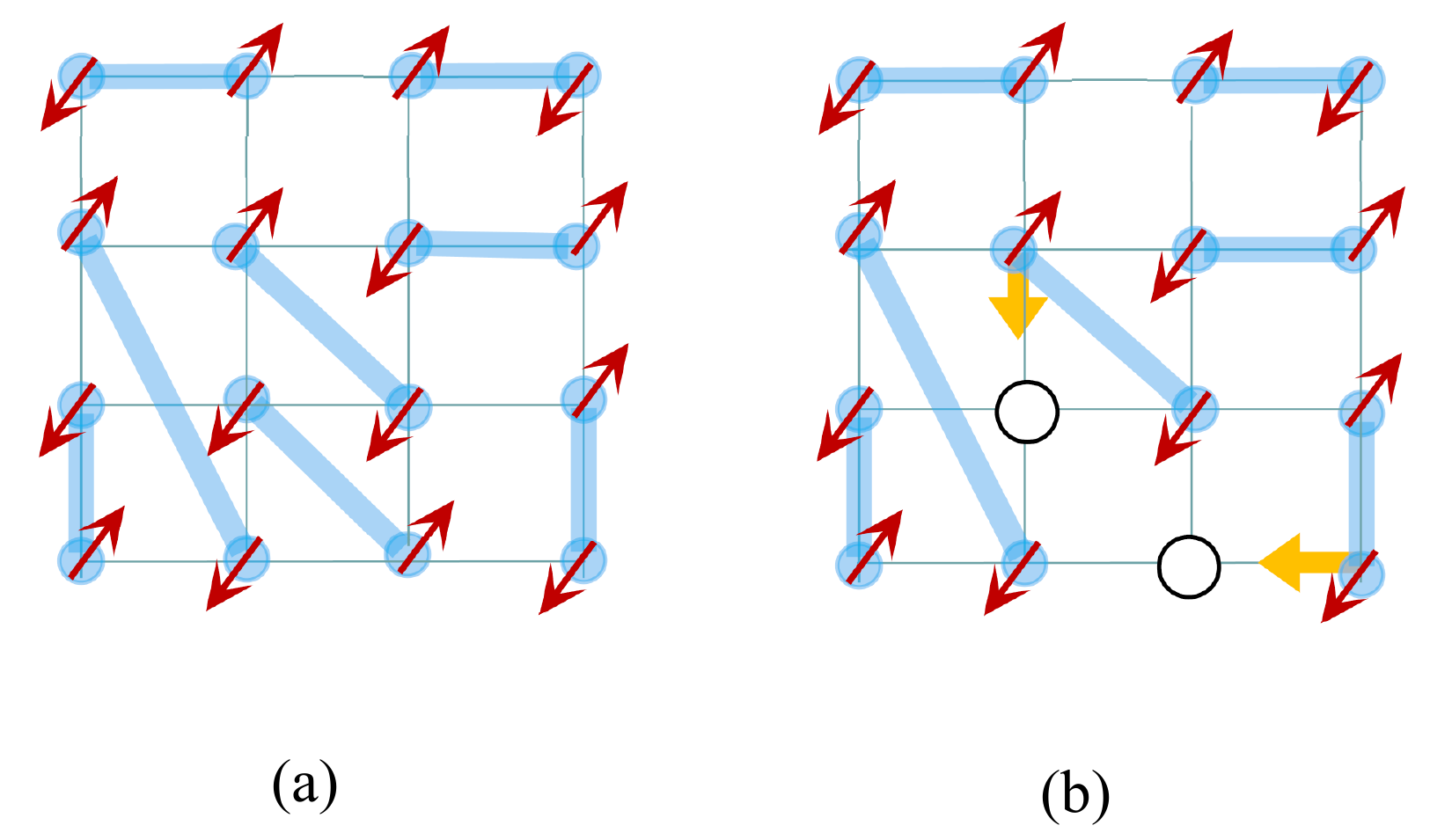} \caption{(Color
online) Schematic illustration of the Gutzwiller projected BCS state
given in Eq. (\protect \ref{BCS}). (a) Half-filling: singlet
electron (Cooper) pairs reduce to the neutral RVB pairs
(blue-colored bonds); (b) Hole doping: the RVB pairs become
partially charged Cooper pairs as they can hop to the hole sites
(e.g., as indicated by yellow arrows).} \label{Fig.1}
\end{figure*}

By contrast, the superconducting ground state obtained in the present work
may be formally written as%
\begin{equation}
|\Psi _{\mathrm{G}}\rangle =\Lambda _{h}\left( \sum_{ij}g_{ij}c_{i\uparrow
}c_{j\downarrow }\right) ^{\frac{N_{h}}{2}}|\mathrm{RVB}\rangle ~
\label{scgs-0}
\end{equation}%
in which $|\mathrm{RVB}\rangle $ denotes a \emph{neutral }spin background
that always remains half-filled as a Mott insulator, whereas the Cooper
pairs\ associated with the doped holes are created by annihilating $N_{h}$
electrons from $|\mathrm{RVB}\rangle $, which are in singlet, d-wave pairing
with an amplitude $g_{ij}$. Apparently such a superconducting state
automatically satisfies the no double occupancy constraint without invoking
the Gutzwiller projection $\hat{P}_{\mathrm{G}}$ as in Eq. (\ref{BCS}). In
particular, it is distinguished from Eq. (\ref{BCS}) by an explicit
separation of the Cooper pairing and RVB pairing at finite doping.
\begin{figure*}[bp]
\centering \includegraphics[width=15cm]{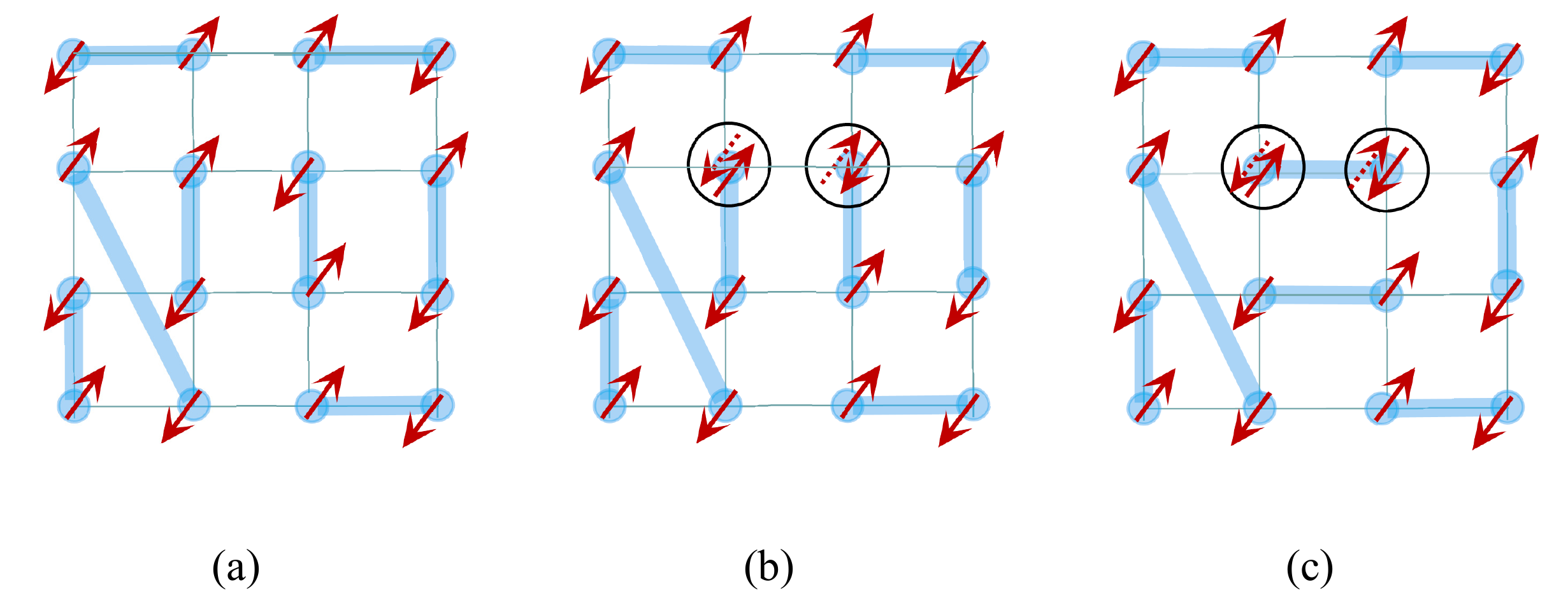}
\caption{(Color online) Schematic illustration of the present ground state (%
\protect \ref{scgs-0}) or (\protect \ref{gsansatz}), which is
structurally
different from the Gutzwiller projected BCS state shown in Fig. \protect \ref%
{Fig.1}. (a) The bosonic RVB pairs in $\left \vert \mathrm{RVB}\right \rangle $%
: each pair only involves spin partners at opposite sublattice
sites; (b) Doped holes are created by annihilating [as indicated by
the dashed arrows which also represent the backflow spinons in the
fractionalization formulation (\protect \ref{gsansatz})] the spins
at the hole sites, and their RVB partners in $\left \vert
\mathrm{RVB}\right \rangle $ automatically become associated with
the doped holes; (c) These partner spins associated with the holes
can also form RVB pairs, which facilitates the Cooper pairing of the
doped holes. Note that the important sign structure of the
wavefunction, i.e., $\Lambda _{h}$ in Eq. (\protect \ref{scgs-0}),
is not directly shown here.} \label{Fig.2}
\end{figure*}

Schematically the RVB pairing in $|\mathrm{RVB}\rangle $ and the Cooper
pairing in $|\Psi _{\mathrm{G}}\rangle $ are illustrated in Figs. 2(a), and
2(b) as well as 2(c), respectively. One easily sees the distinction between
the neutral RVB pairing in Fig. 2(a) and the Cooper pairing in Figs. 2(b)
and 2(c). In the latter, a pair of holes are involved, in which each hole
will be generally associated with a spin via an RVB amplitude [the
blue-colored bond in Fig. 2(b)], since the hole is created by annihilating a
spin whose RVB partner already pre-exists in $|\mathrm{RVB}\rangle $. Note
that the holes are mobile here and thus their spin partners are also
automatically changing with the hopping. Two spin partners associated with a
Cooper pair in Fig. 2(b) can become RVB-paired again to further gain the
superexchange energy, which results in the configuration shown in Fig. 2(c)
and serves the driving force for the Cooper pairing. In this sense, the
superexchange/RVB pairing provides the ultimate pairing \textquotedblleft
glue\textquotedblright \ for superconductivity.

With the Cooper and neutral RVB channels being explicitly differentiated in
Eq. (\ref{scgs-0}), generally some nontrivial phase shift effect will
emerge, as introduced by $\Lambda _{h}$. Specifically, $\Lambda _{h}$ is
given by%
\begin{equation}
\Lambda _{h}\equiv \sum_{\{l_{h}\}}\left( n_{l_{1}}^{h}n_{l_{2}}^{h}\cdot
\cdot \cdot n_{l_{N_{h}}}^{h}\right) \varphi _{h}(l_{1},l_{2},\cdot \cdot
\cdot ,l_{N_{h}})e^{-i\left( \hat{\Omega}_{l_{1}}+\hat{\Omega}_{l_{2}}+\cdot
\cdot \cdot +\hat{\Omega}_{l_{N_{h}}}\right) }  \label{hphih}
\end{equation}%
where $n_{l}^{h}=1-\sum_{\sigma }c_{l\sigma }^{\dagger }c_{l\sigma }\geq 0$
denotes the hole occupation number at site $l$, and $\varphi _{h}$ is a
bosonic wavefunction symmetric with regard to the hole coordinates $%
\{l_{h}\}=l_{1},l_{2},\cdot \cdot \cdot ,l_{N_{h}}$, which is generally
present to ensure gauge invariance of the phase shift fields. Here the phase
shifts $\left \{ \hat{\Omega}_{l_{h}}\right \} ,$ associated with the holes,
will directly act on the \textquotedblleft ghost\textquotedblright \ spin
liquid state $|\mathrm{RVB}\rangle $ to monitor the background spin
correlations. In other words, $\Lambda _{h}$ represents an \textquotedblleft
entanglement\textquotedblright \ between these two channels whose physical
implications and mathematical definition are to be shown below.

\subsubsection{Superconducting phase coherence}

Due to the presence of $\Lambda _{h}$, injecting a hole into the ground
state $|\Psi _{\mathrm{G}}(N_{h})\rangle $ will generally induce a phase
shift by
\begin{equation}
c_{i\sigma }|\Psi _{\mathrm{G}}(N_{h})\rangle \sim e^{i\hat{\Omega}%
_{i}}|\Psi _{\mathrm{G}}(N_{h}+1)\rangle .  \label{pshift}
\end{equation}%
Thus the wavefunction overlap between the bare hole state and the true
ground state of $N_{h}+1$ holes crucially depends on $e^{i\hat{\Omega}_{i}}$.

Furthermore, by noting that the Cooper pairing amplitude already pre-exists
via $g_{ij}$ in Eq. (\ref{scgs-0}), the superconducting
off-diagonal-long-range-order (ODLRO) is essentially determined by (see
below)
\begin{equation}
\left \langle c_{i\uparrow }c_{j\downarrow }\right \rangle \propto \left
\langle \mathrm{RVB}\right \vert e^{i\left( \hat{\Omega}_{i}+\hat{\Omega}%
_{j}\right) }|\mathrm{RVB}\rangle .  \label{phcoh-0}
\end{equation}%
Hence the superconducting phase coherence and the coherence of a Landau (or
more precisely, Bogoliubov) quasiparticle will be simultaneously realized.
In other words, a \textquotedblleft normal state\textquotedblright \
obtained by disordering the phase shift factor $e^{i\hat{\Omega}_{i}}$ will
be intrinsically a non Fermi liquid with a vanishing quasiparticle weight.

In the following we provide a simple proof of Eq. (\ref{phcoh-0}) by taking
\begin{equation}
\varphi _{h}=\text{\textrm{constant}}  \label{hcond}
\end{equation}%
in $\Lambda _{h}$ without loss of generality. Note that Eq. (\ref{scgs-0})
then reduces to
\begin{equation}
|\Psi _{\mathrm{G}}\rangle \propto \hat{D}^{\frac{N_{h}}{2}}|\mathrm{RVB}%
\rangle ~  \label{scgs}
\end{equation}%
with
\begin{equation}
\hat{D}\equiv \sum_{ij}g_{ij}\mathcal{\hat{D}}_{ij}  \label{dd-1}
\end{equation}%
and
\begin{equation}
\mathcal{\hat{D}}_{ij}\equiv e^{-i\left( \hat{\Omega}_{i}+\hat{\Omega}%
_{j}\right) }c_{i\uparrow }c_{j\downarrow }.  \label{ghat}
\end{equation}%
As shown in Appendix A, one has $\left \langle \hat{D}\right \rangle =%
\mathrm{O}(N_{h})$ according to Eq. (\ref{scgs}). Then, so long as $%
g_{ij}\left \langle \mathcal{\hat{D}}_{ij}\right \rangle $ is a short-ranged
function of $\left \vert i-j\right \vert $, an ODLRO can be identified in $%
|\Psi _{\mathrm{G}}\rangle $:

\begin{equation}
g_{ij}\left \langle \mathcal{\hat{D}}_{ij}\right \rangle =\mathrm{O}(\delta )
\label{D-0}
\end{equation}%
with $\delta $ as the doping concentration ($\delta \equiv N_{h}/N$, where $%
N $ denotes the total number of lattice sites). Generally speaking, Eq. (\ref%
{D-0}) represents that the Cooper pairing amplitude is formed (with the
pairing symmetry determined by $g_{ij}$). The true superconducting ODLRO, $%
\left \langle c_{i\uparrow }c_{j\downarrow }\right \rangle $, is thus indeed
determined by the phase coherence condition in Eq. (\ref{phcoh-0}), where $%
\hat{\Omega}_{i}$ sensitively depends on the spin correlation in $|\mathrm{%
RVB}\rangle $.

\subsubsection{Sign structure}

The phase shift $\hat{\Omega}_{i}$ is quantitatively given by%
\begin{equation}
e^{-i\hat{\Omega}_{i}}=e^{-\frac{i}{2}\left( \Phi _{i}^{s}-\Phi
_{i}^{0}\right) }\text{ ,}  \label{phif}
\end{equation}%
in which
\begin{equation}
\Phi _{i}^{s}\equiv \sum_{l\neq i}\theta _{i}(l)\left( \sum_{\sigma }\sigma
n_{l\sigma }^{b}\right) ~,  \label{phis}
\end{equation}%
and%
\begin{equation}
\Phi _{i}^{0}\equiv \sum_{l\neq i}\theta _{i}(l)\text{ ,}~  \label{phi0}
\end{equation}%
where $\theta _{i}(l)=\mathrm{Im}\ln $ $(z_{i}-z_{l})$ ($z_{i}$ is the
complex coordinate of site $i$), and $n_{l\sigma }^{b}$ denotes the spin
occupation number (with index $\sigma )$ at site $l,$ which always satisfies
the single occupancy constraint
\begin{equation}
\text{ }\sum_{\sigma }n_{l\sigma }^{b}=1\text{\ }  \label{constraint=1}
\end{equation}%
acting on the insulating spin state $|\mathrm{RVB}\rangle $.

Then each spin in $|\mathrm{RVB}\rangle $ will contribute to a $\pm \pi $
vortex via $\Phi _{i}^{s}/2$ in Eq. (\ref{phif}) with itself sitting at the
vortex core. Vice versa, each doped hole will be perceived by the spins in $|%
\mathrm{RVB}\rangle $ as introducing a $\pi $ vortex, also via $\Phi
_{i}^{s}/2$, with the hole sitting at the core. It implies that a doped hole
and a neutral spin satisfy a \textquotedblleft mutual semion
statistics\textquotedblright \ as the phase shift $\hat{\Omega}_{i}$ amounts
to giving rise to $\pm \pi $ when one kind of species continuously circles
around the other one once. Note that the single-valueness of Eq. (\ref{phif}%
) will be ensured by combining with $\Phi _{i}^{0}/2$. Thus the total phase
shift added up in $\Lambda _{h}$ represents a nontrivial entanglement
between the doped holes and background spins, which will decide a
\textquotedblleft mutual semion statistics\textquotedblright \ sign
structure in $|\Psi _{\mathrm{G}}\rangle $ that is fundamentally different
from that of a BCS state satisfying the Fermi-Dirac statistics.

One may examine such a sign structure by a thinking experiment in which a
hole in $|\Psi _{\mathrm{G}}\rangle $ goes through a closed loop $c.$ At
each step of nearest-neighbor moving of the hole, a singular phase $0$ or $%
\pi $ is generated via a phase shift $\hat{\Omega}_{i}$ in $\Lambda _{h}$
depending on $\uparrow $ or $\downarrow $ spin that the hole
\textquotedblleft exchanges\textquotedblright \ with. Note that $\Lambda _{h}
$ also produces other phase shift contributed by other spins not
\textquotedblleft exchanged\textquotedblright \ with the hole, but their
effect disappears after counting the total Berry's phase acquired by the
closed-path motion of the hole. In the end, one finds
\begin{equation}
|\Psi _{\mathrm{G}}\rangle \rightarrow (-1)^{N_{h}^{\downarrow }(c)}|\Psi _{%
\mathrm{G}}\rangle  \label{pstring}
\end{equation}%
in which $N_{h}^{\downarrow }(c)$ only counts the total number of the $%
\downarrow $-spins that the hole has \textquotedblleft
exchanged\textquotedblright \ with along the loop $c$ on a square lattice.

The same sign factor $(-1)^{N_{h}^{\downarrow }(c)}$ has been previously
shown to be the precise sign acquired by the hopping of a doped hole through
a closed loop $c$ in the t-J model, i.e., the phase string effect\cite%
{sheng_96,weng_97,WWZ_08}. This effect is proven to be dynamically
irreparable and is thus of statistics nature. So the phase shift $\hat{\Omega%
}$ defined in Eq. (\ref{phif}) is \emph{necessarily} generated by the motion
of a doped hole, representing an emergent new statistics\cite{zaanen_09} in
doped Mott insulators with the Hilbert space restricted by no double
occupancy constraint.

It is noted that at finite doping, the exact topological sign structure
identified based on the t-J model is generally given by\cite{WWZ_08}%
\begin{equation}
\tau _{c}=(-1)^{N_{h}^{\downarrow }(c)}\times (-1)^{N_{h}^{h}(c)}
\label{pstring1}
\end{equation}%
which appears, say, in the partition function
\begin{equation}
Z=\sum_{c}\tau _{c}\mathcal{Z}(c)  \label{partition}
\end{equation}%
where $\mathcal{Z}(c)\geq 0$ for any closed path $c$ of the multi-hole/spin
configurations at arbitrary temperature. Compared to Eq. (\ref{pstring}), an
extra sign factor $(-1)^{N_{h}^{h}(c)}$ appears in Eq. (\ref{pstring1}) in
which $N_{h}^{h}(c)$ counts the number of hole-hole exchanges on the path $c$%
. It is straightforward to verify that that fermionic signs of the doped
holes created by $\left( \sum_{ij}g_{ij}c_{i\uparrow }c_{j\downarrow
}\right) ^{\frac{N_{h}}{2}}$ in Eq. (\ref{scgs-0}) can precisely account for
such a sign factor. Consequently, combined with the phase shift in $\Lambda
_{h}$, the nontrivial sign structure of the t-J model is naturally satisfied
by the ground state $|\Psi _{\mathrm{G}}\rangle $ in Eq. (\ref{scgs-0}) so
long as the neutral spin background $|\mathrm{RVB}\rangle $ does not
contribute to additional statistical signs as shown below.

\subsubsection{$|\mathrm{RVB}\rangle $ as a spin liquid state}

As already mentioned, $|\mathrm{RVB}\rangle $ describes a \textquotedblleft
ghost\textquotedblright \ spin state, which remains one spin at each site of
a square lattice even in the doped case. But the spin state can evolve from
an AFM long-range ordered one to a spin liquid with only short-ranged AFM
correlations as the doping concentration is increased.

Generically $|\mathrm{RVB}\rangle $ can be expressed by%
\begin{equation}
|\mathrm{RVB}\rangle =\sum_{\{ \sigma _{s}\}}\Phi _{\mathrm{RVB}}\left(
\sigma _{1},\sigma _{1},\cdot \cdot \cdot ,\sigma _{N}\right) c_{1\sigma
_{1}}^{\dagger }c_{2\sigma _{2}}^{\dagger }\cdot \cdot \cdot c_{N\sigma
_{N}}^{\dagger }|0\rangle  \label{RVB}
\end{equation}%
in the electron $c$-operator representation. In nature it is a bosonic state
with the wavefunction $\Phi _{\mathrm{RVB}}\left( \{ \sigma _{s}\} \right)
\equiv \sum_{\mathrm{partition}}\prod_{(ij)}(-1)^{i}W_{ij}$ for each given
spin configuration $\{ \sigma _{s}\}=\sigma _{1},\sigma _{1},\cdot \cdot
\cdot ,\sigma _{N}.$ Here the RVB pairing amplitude $(-1)^{i}W_{ij}$
connects two \emph{antiparallel} spins denoted by $i$ (up spin) and $j$
(down spin), with the summation running over all possible pairing partitions
for the given $\{ \sigma _{s}\}$. The staggered sign $(-1)^{i}$ (the
Marshall sign) is explicitly separated from $W_{ij}$ such that the latter
remains a smooth function of the distance between even and odd lattice sites
of a square lattice at different doping concentrations.

Such $|\mathrm{RVB}\rangle $ is a generalized Liang-Docout-Anderson type
bosonic RVB state, which can naturally recover the correct
antiferromagnetism in the zero doping limit\cite{lda_88,WZM_05}. But a
long-range RVB pairing in the antiferromagnetic phase will generally destroy
the phase coherence condition in Eq. (\ref{phcoh-0}) because the $\pm \pi $
vortices carried by the two spinon partners of a long-range RVB pair,
according to Eq. (\ref{phif}), do not compensate each other and result in a
phase disordering. Only can a short-ranged RVB pairing lead to a
vortex-antivortex binding in Eq. (\ref{phcoh-0}) and thus the
superconducting phase coherence. In other words, $|\mathrm{RVB}\rangle $ has
to become a spin liquid in the superconducting phase. How the RVB amplitude $%
W_{ij}$ evolves with doping and self-consistently becomes short-ranged in
the superconducting state will be shown below.

\subsection{Electron fractionalization}

What is the physical implication of the explicit separation of the Cooper
and RVB pairings in the ground state (\ref{scgs-0})? In the following we
show that it actually corresponds to a unique electron fractionalization.

\subsubsection{Ground state in electron fractionalization form}

The ground state $|\Psi _{\mathrm{G}}\rangle $ in Eq. (\ref{scgs-0}) can be
reformulated as a \emph{direct product} state

\begin{equation}
|\Psi _{\mathrm{G}}\rangle =\hat{P}\left( |\Phi _{h}\rangle \otimes |\Phi
_{a}\rangle \otimes |\Phi _{b}\rangle \right) \text{ \ }.  \label{gsansatz}
\end{equation}%
The coefficients, $\varphi _{h}$, $g_{ij}$, and $W_{ij}$, appearing in the
original $|\Psi _{\mathrm{G}}\rangle $ (cf. Sec. II A), are incorporated
into three subsystem states as follows:

\begin{equation}
|\Phi _{h}\rangle \equiv \sum_{\{l_{h}\}}\varphi
_{h}(l_{1},l_{2},...)h_{l_{1}}^{\dagger }h_{l_{2}}^{\dagger }...|0\rangle
_{h}\ ,  \label{bgs}
\end{equation}%
and

\begin{equation}
|\Phi _{a}\rangle \equiv \exp \left( \sum_{ij}\tilde{g}_{ij}a_{i\downarrow
}^{\dagger }a_{j\uparrow }^{\dagger }\right) |0\rangle _{a}~,  \label{phia-0}
\end{equation}%
as well as%
\begin{equation}
|\Phi _{b}\rangle \equiv \exp \left( \sum_{ij}W_{ij}b_{i\uparrow }^{\dagger
}b_{j\downarrow }^{\dagger }\right) |0\rangle _{b}~.  \label{phirvb}
\end{equation}

Here the bosonic wavefunction $\varphi _{h}$ in $|\Phi _{h}\rangle $ defines
a \textquotedblleft holon\textquotedblright \ state with a \emph{bosonic}\
creation operator $h_{l}^{\dagger }$ acting on a vacuum $|0\rangle _{h}$; $%
|\Phi _{b}\rangle $ defines a neutral \textquotedblleft
spinon\textquotedblright \ state with an RVB pairing amplitude $W_{ij}$,
where $b_{i\sigma }^{\dagger }$ as a \emph{bosonic}\ creation operator acts
on a vacuum $|0\rangle _{b}$; and $|\Phi _{a}\rangle $ defines a
\textquotedblleft backflow spinon\textquotedblright \ state with the pairing
amplitude $\tilde{g}_{ij}\equiv \left( -1\right) ^{i}g_{ij}$, where $%
a_{i\sigma }^{\dagger }$ denotes a \emph{fermionic} creation operator acting
on a vacuum $|0\rangle _{a}$.

The projection operator $\hat{P}$ in Eq. (\ref{gsansatz}) is defined by
\begin{equation}
\hat{P}\equiv \hat{P}_{\mathrm{B}}\hat{P}_{s}\text{ ,}  \label{P}
\end{equation}%
in which $\hat{P}_{s}$ will enforce the single-occupancy constraint Eq. (\ref%
{constraint=1}) in the spinon state $|\Phi _{b}\rangle $ such that%
\begin{equation}
|\mathrm{RVB}\rangle \equiv \hat{P}_{s}|\Phi _{b}\rangle  \label{RVB-1}
\end{equation}%
with $n_{i\sigma }^{b}\equiv b_{i\sigma }^{\dagger }b_{i\sigma }$; and $\hat{%
P}_{\mathrm{B}}$ will further enforce
\begin{equation}
\text{ }n_{i\bar{\sigma}}^{a}=n_{i}^{h}n_{i\sigma }^{b}\text{ ,\ }
\label{PB}
\end{equation}%
such that each $a$-spinon always coincides with a holon as $\sum_{\sigma
}n_{i\bar{\sigma}}^{a}=n_{i}^{h}$ according to Eqs. (\ref{PB}) and (\ref%
{constraint=1}) (here $n_{i\bar{\sigma}}^{a}\equiv a_{i\bar{\sigma}%
}^{\dagger }a_{i\bar{\sigma}}$ and $n_{i}^{h}\equiv h_{i}^{\dagger }h_{i}$
with $\bar{\sigma}\equiv -\sigma )$. By applying $\hat{P}$, the physical
Hilbert space is restored in Eq. (\ref{gsansatz}) as schematically shown in
Fig. \ref{Fig.2} in which the $a$-spinons are indicated by the dashed arrows
at the hole sites, while there is always a $b$-spinon indicated by a solid
arrow at each lattice site.

Note that the phase shift factor $e^{-i\hat{\Omega}_{l}}$ in $\Lambda _{h}$
has totally \emph{disappeared}\ in the above direct-product expression Eq. (%
\ref{gsansatz}), where the electrons break up into the fractionalized
building blocks: the holon $h^{\dagger }$, the spinon $b_{\sigma }^{\dagger
} $, and the backflow spinon $a_{\sigma }^{\dagger }$, forming three rather
\textquotedblleft conventional\textquotedblright \ sub-states. The
fractionalization form of the ground state Eq. (\ref{gsansatz}) can be
straightforwardly obtained by substituting into Eq. (\ref{scgs-0}) the
following decomposition form of the electron annihilation operator:
\begin{equation}
c_{i\sigma }=\hat{P}\tilde{c}_{i\sigma }\text{ },  \label{c}
\end{equation}%
with \
\begin{equation}
\tilde{c}_{i\sigma }\equiv h_{i}^{\dagger }a_{i\bar{\sigma}}^{\dagger
}(-\sigma )^{i}e^{i\hat{\Omega}_{i}}\text{ },  \label{decomp}
\end{equation}%
which acts on the insulating \textquotedblleft vacuum\textquotedblright \ $%
|0\rangle _{h}\otimes |0\rangle _{a}\otimes |\mathrm{RVB}\rangle $. On the
other hand, in the neutral spin state $|\mathrm{RVB}\rangle $, the $%
c^{\dagger }$-operator can be reexpressed in terms of the bosonic spinon
operator $b_{i\sigma }^{\dagger }$ to result\cite{WZM_05} in Eq. (\ref{RVB-1}%
) from Eq. (\ref{RVB}), according to the decomposition Eq. (\ref{mutual})
given in Sec. III, where it is further demonstrated that the full spin
operator can be expressed as%
\begin{equation}
\mathbf{S}_{i}=\hat{P}\mathbf{\tilde{S}}_{i}\text{ ,}  \label{s1}
\end{equation}%
with

\begin{equation}
\mathbf{\tilde{S}}_{i}\equiv \mathbf{S}_{i}^{b}+\mathbf{S}_{i}^{a}\text{ ,}
\label{s-op}
\end{equation}%
where $\mathbf{S}_{i}^{b}$ denotes the spin operators for $b$-spinons
[defined in Eqs. (\ref{sbz}) and (\ref{sb+})] and $\mathbf{S}_{i}^{a}$ for $%
a $-spinons [defined in Eqs. (\ref{saz}) and (\ref{sa+})].

\subsubsection{Effective Hamiltonian}

The electron fractionalization form [Eq. (\ref{gsansatz})] of the ground
state $|\Psi _{\mathrm{G}}\rangle $ will make the manipulation of the
superconducting state more easily than in the original form [Eq. (\ref%
{scgs-0})].

Define

\begin{equation}
|\tilde{\Psi}_{\mathrm{G}}\rangle \equiv |\Phi _{h}\rangle \otimes |\Phi
_{b}\rangle \otimes |\Phi _{a}\rangle \text{ ,}  \label{gsansatz1}
\end{equation}%
such that $|\Psi _{\mathrm{G}}\rangle =\hat{P}|\tilde{\Psi}_{\mathrm{G}%
}\rangle $. Then $c_{i\sigma }|\Psi _{\mathrm{G}}\rangle =\hat{P}\tilde{c}%
_{i\sigma }|\tilde{\Psi}_{\mathrm{G}}\rangle $ and $\mathbf{S}_{i}|\Psi _{%
\mathrm{G}}\rangle =\hat{P}\mathbf{\tilde{S}}_{i}|\tilde{\Psi}_{\mathrm{G}%
}\rangle $, in which $\tilde{c}_{i\sigma }$ and $\mathbf{\tilde{S}}_{i}$
directly act on the fractionalized states.

Based on the t-J model, we find that the direct product state $|\tilde{\Psi}%
_{\mathrm{G}}\rangle $ in Eq. (\ref{gsansatz1}) can be effectively
determined as the \emph{ground state} of the following effective Hamiltonian%
\begin{equation}
H_{\mathrm{eff}}=H_{h}+H_{s}+H_{a}\text{ ,}  \label{heff}
\end{equation}%
which is composed of a holon hopping term
\begin{equation}
H_{h}=-t_{h}\sum_{\langle ij\rangle }\left( e^{iA_{ij}^{s}}\right)
h_{i}^{\dagger }h_{j}+h.c.,  \label{hh}
\end{equation}%
a $b$-spinon pairing term
\begin{equation}
H_{s}=-J_{s}\sum_{\langle ij\rangle \sigma }\left( e^{i\sigma
A_{ij}^{h}}\right) b_{i\sigma }^{\dagger }b_{j-\sigma }^{\dagger }+h.c.,
\label{hs}
\end{equation}%
and an $a$-spinon term%
\begin{eqnarray}
H_{a} &=&-t_{a}\sum_{\left \langle ij\right \rangle \sigma }e^{-i\phi
_{ij}^{0}}a_{i\sigma }^{\dagger }a_{j\sigma }-J_{a}\sum_{\langle ij\rangle
}\eta _{ij}\hat{\Delta}_{ij}^{a}+{h.c.}  \notag \\
&&+J\sum_{\langle ij\rangle }\left( \mathbf{S}_{i}^{a}\cdot \mathbf{S}%
_{j}^{b}+\mathbf{S}_{i}^{b}\cdot \mathbf{S}_{i}^{a}\right) \text{ ,}
\label{ha}
\end{eqnarray}%
where $\hat{\Delta}_{ij}^{a}\equiv \sum_{\sigma }\sigma a_{i\sigma
}^{\dagger }a_{j-\sigma }^{\dagger }$ and $\eta _{ij}=+(-)$ for $j=i\pm \hat{%
x}(\hat{y})$ is a d-wave sign factor. Note that the chemical (Lagrangian
multiplier) terms implementing
\begin{equation}
\sum_{i}h_{i}^{\dagger }h_{i}=\sum_{i\sigma }a_{i\sigma }^{\dagger
}a_{i\sigma }=\delta N
\end{equation}%
and%
\begin{equation}
\sum_{i\sigma }b_{i\sigma }^{\dagger }b_{i\sigma }=N
\end{equation}%
are all omitted in Eqs. (\ref{hh}), (\ref{hs}), and (\ref{ha}) for
simplicity. One can always add them back in real calculations.

Based on Eq. (\ref{gsansatz1}), the parameters, $t_{h}\sim t$, $J_{s}\sim J$%
, $t_{a}\sim t$, and $J_{a}\sim J\left \vert \left \langle \hat{\Delta}%
^{a}\right \rangle \right \vert $ in Eqs. (\ref{hh})-(\ref{ha}) can be
determined as \emph{variational} parameters minimizing the ground state
energy of $\hat{P}|\tilde{\Psi}_{\mathrm{G}}\rangle $ at a given doping
concentration, which will involve the projection $\hat{P}$ and whose
detailed magnitudes will not affect the general consequences to be outlined
in the next subsection.

In $H_{h}$ and $H_{s}$, the $h$-holons and $b$-spinons are generally coupled
to the \textrm{U(1)}$\otimes \mathrm{U(1)}$ gauge fields, $A_{ij}^{s}$ and $%
A_{ij}^{h},$ respectively, in Eqs. (\ref{hh}) and (\ref{hs}), which are
topological (mutual Chern-Simons) fields as their gauge-invariant flux
strengths in an arbitrary counter-clockwise closed loop $c$ are constrained
to the numbers of spinon and holon matter fields within the enclosed area $%
\Sigma _{c}$, respectively,%
\begin{equation}
\sum_{c}A_{ij}^{s}=\pi \sum_{l\in \Sigma _{c}}\left( n_{l\uparrow
}^{b}-n_{l\downarrow }^{b}\right) ,  \label{cond1}
\end{equation}%
and
\begin{equation}
\sum_{c}A_{ij}^{h}=\pi \sum_{l\in \Sigma _{c}}n_{l}^{h}.  \label{cond2}
\end{equation}%
The link variables, $A_{ij}^{s}$ and $A_{ij}^{h}$, can be regarded as
mediating the mutual statistics coupling between the charge and spin degrees
of freedom, i.e., the \textquotedblleft mutual semion
statistics\textquotedblright \ entanglement\ introduced by the phase shift
factor $e^{-i\hat{\Omega}_{i}}$ in the original ground state Eq. (\ref%
{scgs-0}). In addition, the constant link field $\phi _{ij}^{0}$ in Eq. (\ref%
{ha}) describes a non-dynamic $\pi $ flux per plaquette, which is originated
from $\Phi _{i}^{0}$ term in Eq. (\ref{phif}).

\subsubsection{Ground state as a mean-field solution}

Then the ground state (\ref{gsansatz1}) as a self-consistent mean-field
solution of $H_{\mathrm{eff}}$ in Eq. (\ref{heff}) can be constructed as
follows.

First of all, \emph{suppose} the holon state $|\Phi _{h}\rangle $ governed
by $H_{h}$ in Eq. (\ref{hh}) become Bose-condensed [cf. Eq. (\ref{hcond})].
Such holon condensation will then lead to $A_{ij}^{h}\longrightarrow \bar{A}%
_{ij}^{h}$, with $\bar{A}_{ij}^{h}$ depicting a uniform flux $\sum_{{\large %
\Box }}\bar{A}_{ij}^{h}=\pi \delta $ per plaquette, in terms of Eq. (\ref%
{cond2}).

Then $H_{s}$ in Eq. (\ref{hs}) can be diagonalized, resulting in a
mean-field solution $|\Phi _{b}\rangle $ given in Eq. (\ref{phirvb}), in
which $W_{ij}$ $=0$ if both $i$ and $j$ belong to the same sublattice and
decays exponentially at large spatial separations for opposite sublattice
sites $i$ and $j$: $\left \vert W_{ij}\right \vert \propto e^{-\frac{|%
\mathbf{r}_{ij}|^{2}}{2\xi ^{2}}}$\cite{WZM_05}. Here $\mathbf{r}_{ij}$ is
the spatial distance and $\xi $ is the characteristic pair size determined
by the doping concentration: $\xi =a\sqrt{\frac{2}{\pi \delta }}$ ($a$ is
the lattice constant). Hence, the spin background $|\Phi _{b}\rangle $
indeed becomes \emph{short-ranged} at finite doping with a finite spin gap $%
E_{g}\propto \delta J$\cite{weng_07,WZM_05}. Once the $b$-spinons are all
short-range paired up in $|\Phi _{b}\rangle $, the fluctuations of $%
A_{ij}^{s}$ would become negligible for the long-wavelength physics, i.e., $%
\sum_{c}A_{ij}^{s}\approx 0$ for a large loop $c$ as compared to $\xi $,
according to Eq. (\ref{cond1}). Self-consistently, the two subsystems of the
holons and $b$-spinons are decoupled as depicted by $|\Phi _{h}\rangle
\otimes $ $|\Phi _{b}\rangle ,$ as the ground state of
\begin{equation}
H_{\mathrm{string}}=H_{h}+H_{s},  \label{psmodel}
\end{equation}
which is known as the phase string model\cite{weng_99,weng_07} or mutual
Chern-Simons gauge theory model\cite{YTQW_11}.

It is interesting to point out that $|\mathrm{RVB}\rangle =\hat{P}_{s}|\Phi
_{b}\rangle $ is of the same form as the Liang-Docout-Anderson type RVB
wavefunction at half-filling, which has been previously proposed\cite{lda_88}
as a very accurate variational ground-state wavefunction for the Heisenberg
model. Indeed, at half-filling, in the absence of holes, $|\Psi _{\mathrm{G}%
}\rangle $ simply reduces to $|\mathrm{RVB}\rangle $, with $\xi \rightarrow
\infty $ or $W_{ij}$ obeying the power law at large spatial separation of $%
ij $: $\left \vert W_{ij}\right \vert \propto 1/\left \vert \mathbf{r}%
_{ij}\right \vert ^{3}$\cite{WZM_05}. The ground state energy and staggered
magnetization of the antiferromagnetic ordering determined numerically based
on such $|\mathrm{RVB}\rangle ,$ obtained from $H_{s},$ are highly accurate
as compared to the exact numerical results\cite{WZM_05}. This indicates that
the bosonic RVB mean-field description in Eq. (\ref{hs}), which reduces to
the Schwinger-boson mean-field theory at half-filling\cite{sfermion}, has
accurately captured both short-range and long-wavelength correlations of the
Heisenberg model in this limit.

Finally, note that the fermionic backflow $a$-spinons in $H_{a}$ [Eq. (\ref%
{ha})] are \emph{gauge neutral} without coupling to the internal mutual
Chern-Simons gauge fields. Here the last scattering term with the $b$%
-spinons in Eq. (\ref{ha}) can be safely omitted in determining the ground
state $|\Phi _{a}\rangle $, due to the above-mentioned gap $E_{g}$ opening
up in the spin excitation involving $b$-spinons in $|\Phi _{b}\rangle $.
Then the bilinear terms in Eq. (\ref{ha}) can be straightforwardly
diagonalized with a proper gauge choice of $\phi _{ij}^{0}$, to result in
Eq. (\ref{phia-0}) with a d-wave amplitude $\tilde{g}_{ij}$. Due to the
presence of $\phi _{ij}^{0}$, contributing to a $\pi $-flux per plaquette,
the $a$-spinons will form Fermi pockets at both $(0$,$0)$ and antinodal
point $(\pi $,$0),$ etc., and a staggered current loop is expected to be
present at $\Delta _{ij}^{a}\neq 0$. The physical implications of the $a$%
-spinon excitations will be further discussed later.

To end this section, let us examine the electron pairing parameter $\Delta
_{ij}^{\mathrm{SC}}\equiv \left \langle c_{i\uparrow }c_{j\downarrow }\right
\rangle $ based on Eq. (\ref{decomp}). It can be expressed in the present
fractionalized state by%
\begin{equation}
\Delta _{ij}^{\mathrm{SC}}\propto \Delta _{ij}^{a}\left \langle e^{i\left(
\hat{\Omega}_{i}+\hat{\Omega}_{i}\right) }\right \rangle \text{.}
\label{scodlro}
\end{equation}%
Namely, the pairing amplitude and symmetry will be determined by the pairing
order parameter of the $a$-spinons, and the superconducting phase coherence
is decided by Eq. (\ref{phcoh-0}). The latter is realized as the RVB pairing
of the $b$-spinons in $|\Phi _{b}\rangle $ becomes short-ranged with a
finite\ $\xi $, such that the $\pi $-vortices and -antivortices attached to
them, according to Eq. (\ref{phis}), are all confined to form the
vortex-antivortex pairs. Namely, superconductivity will be protected by a
\textquotedblleft ghost\textquotedblright \ spin liquid state. As noted
before, the superconducting phase coherence can either disappear as $\xi
\rightarrow \infty $ either in the long-range AFM state near half-filling,
or in the overdoped regime when the RVB pairing in $|\mathrm{RVB}\rangle $
is diminished by doping.

\subsection{Elementary excitations}

Once the ground state ansatz Eq. (\ref{scgs-0}) or its fractionalization
form Eq. (\ref{gsansatz}) is determined, the corresponding low-lying
elementary excitations, which reflect the novel correlations in the ground
state, will also naturally manifest.

In fact, the effective Hamiltonian Eq. (\ref{heff}) determines not only the
ground state $|\tilde{\Psi}_{\mathrm{G}}\rangle $ in Eq. (\ref{gsansatz1}),
but also some nontrivial excited states. In the following we first show the
existence of two novel elementary excitations which are uniquely governed by
$H_{\mathrm{eff}}$. Then we show that a conventional Bogoliubov
quasiparticle excitation will also appear as a \emph{collective mode} that
goes beyond $H_{\mathrm{eff}},$ which will remain protected within the basic
characteristic energy scale $E_{g}$ in this non-BCS superconducting state.

\subsubsection{Spin-roton excitations}

In the ground state Eq. (\ref{gsansatz}), one does not see the trace of the
electrons directly -- such a strongly correlated electron system seems
entirely \emph{fractionalized,} as described by the bosonic RVB paired
spinons, Bose condensed holons, and d-wave paired backflow spinons, which
form a direct product (a generalized \textquotedblleft spin-charge
separation\textquotedblright ) state.

However, we find that the single $b$-spinons and holons will not be truly
present in the low-lying energy spectrum to become real elementary
excitations. This is because $b$-spinons and holons are not gauge neutral --
they carry the \textquotedblleft gauge charges\textquotedblright \ of the
mutual Chern-Simons fields, $A^{h}$ and $A^{s}$, while provide the
\textquotedblleft topological sources\textquotedblright \ to generate $A^{s}$
and $A^{h}$, respectively, and thus their excitations, by breaking up the
\emph{correlated patterns} formed in the ground state, will generally invite
nonlocal responses from the whole system, which would make such excitations
too costly\cite{ZMW_03}.

For instance, one can imagine spinon excitations created by breaking up an
RVB pair in $|\Phi _{b}\rangle $, described by the effective Hamiltonian $%
H_{s}$ in Eq. (\ref{hs}). However, each unpaired $b$-spinon will induce
vortexlike superfluid currents via $A^{s}$ from the condensed holons
according to $H_{h}$ in Eq. (\ref{hh}), leading to the so-called \emph{%
spinon-vortex} composite object which is logarithmically divergent in
energy, as discussed in Refs. \cite{MW_02,WM_02,WQ_06}. Therefore, these
spinons can only exist in the RVB pair condensate in the ground state, where
such vortex currents get effectively cancelled out due to Eq. (\ref{cond1}),
but not as a single excitation at low energy. Namely, single $b$-spinon
excitations will be \textquotedblleft confined\textquotedblright \cite%
{YTQW_11}\ in the bulk of the superconductor.

On the other hand, an integer ($S=0$ and $1$) spin excitation involving a
bound pair of spinons excited in $|\Phi _{b}\rangle $ are still allowed\cite%
{WQ_06,MW_10}, in which the effect of vortex and antivortex bound to
individual spinons get cancelled out in the long distance in $H_{h}$ such
that its excitation energy becomes finite with a mean-field gap $E_{g}$
according to $H_{s}$.

Such a neutral spin mode carrying an either integer $S=0$ or $1$ quantum
number, is called a spin-roton\cite{MW_10}. The spin-roton excitations will
not destroy the phase coherence at finite temperature until $T_{c}$, where
the spin-rotons disassociate into free spinon-vortices\cite%
{MW_02,WM_02,WQ_06}. It has been shown\cite{MW_10} that a simple $T_{c}$
formula: \
\begin{equation}
T_{c}\simeq \frac{E_{g}}{6k_{\mathrm{B}}}  \label{tc}
\end{equation}%
can be determined with $E_{g}\sim \delta J$ denoting the core energy of the
spin-rotons, degenerate for $S=0$ and $1$, which is in excellent agreement
with the experiments, with the \textquotedblleft
resonancelike\textquotedblright \ modes observed in the Raman $A_{\mathrm{1g}%
} $ channel and neutron scattering measurements consistently interpreted as
the spin-roton excitations with $S=0$ and $1$, respectively\cite{Uemura}. It
also provides a natural explanation why the two modes are energetically
degenerate in the experiment\cite{Uemura}.

Here the spin-rotons are the most essential elementary excitations of
non-BCS-type above the superconducting ground state Eq. (\ref{scgs-0}) or
Eq. (\ref{gsansatz}), which directly controls the superconducting phase
coherence Eq. (\ref{phcoh-0}) via a characteristic energy $E_{g}$. In the
AFM long-range ordered state near half-filling, one has $E_{g}\rightarrow 0$
such that $T_{c}$ vanishes, and the $S=1$ spin-roton excitation will
naturally reduce to the gapless spin wave.

\subsubsection{Fermionic $a$-spinon excitation}

In the ground state Eq. (\ref{gsansatz}), there are two distinct branches of
spinons. The $S=1$ spin-roton excitations related to the dynamic correlation
function of $\mathbf{S}_{i}^{b}$ have been discussed above. According to Eq.
(\ref{s-op}), the backflow $a$-spinons will contribute to another branch of $%
S=1$ excitations as governed by $H_{a}$ in Eq. (\ref{ha}). Here the single $%
a $-spinons are gauge-neutral and can be excited by breaking up the d-wave
pairs in $|\Phi _{a}\rangle $ [Eq. (\ref{phia-0})], which is expected to
provide a characteristically different spectral contribution \emph{below}
the spin-roton \textquotedblleft resonance\textquotedblright \ modes
mentioned above [at $E>$ $E_{g}$, such an $S=1$ mode composed of the $a$%
-spinons may strongly decay into the spin-roton mode via the last term in
Eq. (\ref{ha})].

Besides contributing to the spin spectral function, a single $a$-spinon can
also directly appear in the single-particle channel. According to the
decomposition Eq. (\ref{decomp}), a coherent term may emerge as the first
term in

\begin{equation}
\tilde{c}_{i\sigma }=h_{0}^{\ast }a_{i\bar{\sigma}}^{\dagger }(-\sigma
)^{i}e^{i\hat{\Omega}_{i}}+:h_{i}^{\dagger }:a_{i\bar{\sigma}}^{\dagger
}(-\sigma )^{i}e^{i\hat{\Omega}_{i}}  \label{c3}
\end{equation}%
with the holon condensation $\left \langle h_{i}^{\dagger }\right \rangle
=h_{0}^{\ast }$ ($:h_{i}^{\dagger }:\equiv h_{i}^{\dagger }-h_{0}^{\ast }$)
and the superconducting phase coherence $\left \langle e^{i\hat{\Omega}%
_{i}}\right \rangle \neq 0$. In other words, the $a$-spinon excitation may
be directly probed by ARPES as a coherent term appearing below $T_{c}$ with
a weight $\left \vert h_{0}\right \vert ^{2}\propto \delta $ and
disappearing above $T_{c}$ when $\left \langle e^{i\hat{\Omega}%
_{i}}\right
\rangle =0$. Such a coherent term will be nonetheless
distinguished from the true quasiparticle excitation, which will be obtained
as a \emph{collective mode}, i.e., a bound state of the holon and $a$-spinon
by the singular phase shift $\hat{\Omega}_{i}$ from the second term in Eq. (%
\ref{c3}), as to be given in the following subsection. Since the low-lying $%
a $-spinon excitation will appear near the antinodal region [i.e., the
momentum $(0,\pi )$, etc.], while the true quasiparticle, after absorbing
the phase shift field $e^{i\hat{\Omega}_{i}}$, is a nodal quasiparticle
around $(\pi /2$, $\pi /2$)$,$ there is a \textquotedblleft
dichotomy\textquotedblright \ between these two kinds of excitations in the
single-particle channel, which will be explored in detail elsewhere.

Finally, it is noted that in a strong magnetic field, the d-wave pairing of
the $a$-spinons may be first broken down by the Zeeman energy \emph{before}
the spin-rotons (of energy $\gtrsim E_{g}$) in $|\mathrm{RVB}\rangle $ get
excited to destroy the phase coherence in Eq. (\ref{phcoh-0}). In this case,
the $a$-spinons in Eq. (\ref{ha}) will form coherent Fermi pockets with $%
\Delta _{ij}^{a}=0$ such that the superconducting order parameter in Eq. (%
\ref{scodlro}) can also vanish. Then a new normal state characterized by
small Fermi pockets of the $a$-spinons can be realized by applying a
sufficiently strong magnetic field without encountering the phase
disordering boundary. Of course, such a $T=0$ transition caused by the
Zeeman effect should be compared to another possible route to a normal state
via the generation of new type of magnetic vortices at strong magnetic fields%
\cite{MW_02,WM_02,WQ_06}.

\subsubsection{Quasiparticle as a collective mode}

So far we have discussed two novel elementary excitations in the
superconducting state, i.e., the spin-rotons and the $a$-spinons, which are
apparently non-BCS-like as determined by $H_{\mathrm{eff}}$ in Eq. (\ref%
{heff}). In the following we point out that the conventional Bogoliubov
quasiparticle excitation will reemerge as a \emph{collective mode }in the
full t-J Hamiltonian, although it is not an eigen solution of $H_{\mathrm{eff%
}}$. In other words, a Bogoliubov quasiparticle can be regarded as a bound
state of the fractionalized building blocks, which are glued by the residual
interaction in the original t-J model\cite{WST_00,ZMW_03}. However, such a
quasiparticle excitation will be stable only in a sufficiently long-range,
low-energy regime, where the superconducting state will still behave like a
conventional d-wave BCS superconductor as other exotic modes are not yet
excited.

In order to generally trace a quasiparticle excitation, one may directly
create a bare hole (particle) by the electron $c$-operator on the ground
state $|\Psi _{\mathrm{G}}\rangle $, and then follow its behavior via the
following equation-of-motion (cf. Sec. III D for details):\  \  \  \  \  \  \  \  \
\  \  \  \  \  \  \  \
\begin{equation}
-i\partial _{t}c_{\mathbf{k}\sigma }\simeq -(\epsilon _{\mathbf{k}}-\mu )c_{%
\mathbf{k}\sigma }-\Delta _{\mathbf{k}}\sigma c{_{-\mathbf{k}-\sigma
}^{\dagger }}+\mathrm{decay}\text{ \textrm{term}}+\mathrm{scattering}\text{
\textrm{term}}  \label{eqom}
\end{equation}%
which is obtained after a linearization in terms of the mean-field order
parameters. Here $\Delta _{\mathbf{k}}\propto \Delta _{\mathbf{k}}^{SC}$ is
given by Eq. (\ref{G}). The decay term is given in Eq. (\ref{decay term})
which corresponds to the process discussed in the above subsection that a
doped hole dissolves into an $a$-spinon, i.e.,
\begin{equation}
\mathrm{decay}\text{ \textrm{term }}\sim \left \langle e^{i\hat{\Omega}%
_{i}}\right \rangle h_{0}^{\ast }a_{\mathbf{k}\bar{\sigma}}^{\dagger }
\label{decay}
\end{equation}%
in the superconducting background of $h_{0}^{\ast }\neq 0$ and $%
\left
\langle e^{i\hat{\Omega}_{i}}\right \rangle \neq 0$. Such a coherent
mode will appear in the antinodal region where the Fermi pockets of the $a$%
-spinons locate, and disappear once the superconducting phase coherence gets
lost at $\left \langle e^{i\hat{\Omega}_{i}}\right \rangle =0$. The
scattering term reads [cf. Eq. (\ref{scatt})]%
\begin{equation}
\mathrm{scattering}\text{ }\mathrm{term}=\sum_{\mathbf{q}}\left( 2t\Gamma _{%
\mathbf{k+q}}-J\Gamma _{\mathbf{q}}\right) \left[ \sigma c_{\mathbf{k+q}%
\sigma }S_{\mathbf{q}}^{bz}+c_{\mathbf{k+q-}\sigma }S_{\mathbf{q}}^{b-\sigma
}\right]  \label{scatter}
\end{equation}%
with $\Gamma _{\mathbf{k}}\equiv \cos k_{x}a+\cos k_{y}a$, which represents
a process that the doped hole scatters with the spin-rotons excitations
created by $\mathbf{S}_{i}^{b}$ above the resonancelike energy $E_{g}$.

Therefore, if we only focus on the low-lying excitation below the spin-roton
energy $E_{g}$ and around the \emph{nodal} region, the scattering and decay
terms in Eq. (\ref{eqom}) can be all neglected. Then Eq. (\ref{eqom}) and
its quasiparticle counterpart can be combined to give rise to an elementary
excitation of the Bogoliubov quasiparticle type $\alpha _{\mathbf{k}\sigma
}^{\dagger }\propto u_{\mathbf{k}}c_{\mathbf{k}\sigma }^{\dagger }+\sigma v_{%
\mathbf{k}}c_{-\mathbf{k-}\sigma }$, which leads to
\begin{equation}
-i\partial _{t}\alpha _{\mathbf{k}\sigma }^{\dagger }|\Psi _{\mathrm{G}%
}\rangle =E_{\mathbf{k}}\alpha _{\mathbf{k}\sigma }^{\dagger }|\Psi _{%
\mathrm{G}}\rangle
\end{equation}%
and $\alpha _{\mathbf{k}\sigma }|\Psi _{\mathrm{G}}\rangle =0$, with $u_{%
\mathbf{k}}$, $v_{\mathbf{k}}$, and the energy spectrum $E_{\mathbf{k}}=%
\sqrt{(\epsilon _{\mathbf{k}}-\mu )^{2}+\left( \Delta _{\mathbf{k}}\right)
^{2}}$ given in Sec. III D, not different from a usual d-wave nodal
quasiparticle in a BCS framework. Here the chemical potential $\mu $ is
determined by requiring $\sum_{\mathbf{k\sigma }}c{_{\mathbf{k}\sigma
}^{\dagger }}c_{\mathbf{k}\sigma }=(1-\delta )N$, and one arrives at a very
important conclusion that a d-wave Bogoliubov quasiparticle excitation is
still well preserved in the present non-BCS-like superconducting state. It
is not given by the effective Hamiltonian $H_{\mathrm{eff}}$, but emerges as
a collective mode of the original t-J model, which is ensured by the
superconducting ODLRO (\ref{scodlro}) protected by a finite minimal
spin-roton energy $E_{g}$.\  \  \  \

\section{Microscopic Justification}

In this section, we justify the ground state ansatz, Eqs. (\ref{scgs-0}) and
(\ref{gsansatz}), as well as the elementary excitations outlined in Sec. II,
based on the t-J model. We shall start with an exact reformulation of the
t-J model in an all-boson formalism in which the hidden sign structure can
be explicitly revealed. Such a precise sign structure will then play an
essential role in determining the peculiar structure of the ground state and
elementary excitations in terms of the electron fractionalization.

\subsection{Phase string representation of the t-J model}

The t-J Hamiltonian is a minimal model of doped Mott insulators, with the
Hilbert space restricted by the no double occupancy constraint (\ref%
{mottness}) in the hole-doped case. It has been rigorously demonstrated that
the fermion signs are completely diminished at half-filling due to the
\textquotedblleft Mottness\textquotedblright \ enforced by the constraint (%
\ref{mottness}), where the t-J model reduces to the AFM Heisenberg model.
The residual fermion signs only start to \emph{re-emerge} upon doping, which
can be mathematically described by the so-called phase string effect\cite%
{sheng_96,weng_97,WWZ_08}.

In order to explicitly keep track of such a sign structure, the phase string
representation of the t-J model has been previously introduced\cite{weng_97}%
, in which the electron annihilation operator can be fully bosonized by the
following decomposition
\begin{equation}
c_{i\sigma }=h_{i}^{\dagger }b_{i\sigma }e^{i\hat{\Theta}_{i\sigma }}
\label{mutual}
\end{equation}%
in terms of the bosonic holon creation operator $h_{i}^{\dagger }$ and the
bosonic spinon annihilation operator $b_{i\sigma }$, with the phase string
effect explicitly embedded in the phase factor%
\begin{equation}
e^{i\hat{\Theta}_{i\sigma }}\equiv (-\sigma )^{i}e^{i\frac{1}{2}\left[ \Phi
_{i}^{s}-\Phi _{i}^{0}-\sigma \Phi _{i}^{h}\right] }  \label{psfactor}
\end{equation}%
In $e^{i\hat{\Theta}_{i\sigma }}$, $\Phi _{i}^{s}$ and $\Phi _{i}^{0}$ are
defined in Eqs. (\ref{phis}) and (\ref{phi0}), respectively, and $\Phi
_{i}^{h}$ is given as follows%
\begin{equation}
\Phi _{i}^{h}=\sum_{l\neq i}\theta _{i}(l)n_{l}^{h}  \label{phih}
\end{equation}%
which is nonlocally associated with the holon occupation number $n_{l}^{h}$
at site $l$. Here $e^{i\hat{\Theta}_{i\sigma }}$ also plays a role to
restore the fermionic statistics of $c_{i\sigma }$, in the Hilbert space
restricted by the single occupancy constraint
\begin{equation}
\sum_{\sigma }n_{i\sigma }^{b}+n_{i}^{h}=1.  \label{single}
\end{equation}%
Correspondingly the spin operators $\mathbf{S}_{i}=\mathbf{S}_{i}^{b}$ with $%
\mathbf{S}_{i}^{b}$ are defined by
\begin{equation}
S_{i}^{bz}\equiv \frac{1}{2}\sum_{\sigma }\sigma b_{i\sigma }^{\dagger
}b_{i\sigma }  \label{sbz}
\end{equation}%
and%
\begin{equation}
S_{i}^{b+}\equiv (-1)^{i}b_{i\uparrow }^{\dagger }b_{i\downarrow }e^{i\Phi
_{i}^{h}}\text{, \  \  \  \  \  \ }S_{i}^{b-}\equiv (-1)^{i}b_{i\downarrow
}^{\dagger }b_{i\uparrow }e^{-i\Phi _{i}^{h}}  \label{sb+}
\end{equation}%
which involve the holon degree of freedom via $\Phi _{i}^{h}$ defined above.

Then the t-J Hamiltonian
\begin{eqnarray*}
H_{\mathrm{t-J}} &=&H_{t}+H_{J} \\
&\equiv &-t\sum_{\langle ij\rangle \sigma }c_{i\sigma }^{\dagger }c_{j\sigma
}+h.c.~+J\sum_{\langle ij\rangle }\left( \mathbf{S}_{i}\cdot \mathbf{S}_{j}-%
\frac{n_{i}n_{j}}{4}\right)
\end{eqnarray*}%
under the constraint $n_{i}\equiv \sum_{\sigma }c_{i\sigma }^{\dagger
}c_{i\sigma }\leq 1$ can be reformulated as\cite{weng_97}
\begin{equation}
H_{t}=-t\sum_{\langle ij\rangle \sigma }\left( h_{i}^{\dagger
}h_{j}e^{iA_{ij}^{s}}\right) \left( b_{j\sigma }^{\dagger }b_{i\sigma
}e^{i\phi _{ji}^{0}+i\sigma A_{ji}^{h}}\right) +h.c.~  \label{ht}
\end{equation}%
and%
\begin{equation}
H_{J}=-\frac{J}{2}\sum_{\langle ij\rangle }\left( \hat{\Delta}%
_{ij}^{s}\right) ^{\dagger }\hat{\Delta}_{ij}^{s}~,  \label{hj}
\end{equation}%
with the bosonic RVB order operator
\begin{equation}
\hat{\Delta}_{ij}^{s}=\sum_{\sigma }e^{-i\sigma A_{ij}^{h}}b_{i\sigma
}b_{j-\sigma }~  \label{ds-0}
\end{equation}%
for the NN sites.

Here the unique feature is the presence of three link fields: $A_{ij}^{s}$, $%
A_{ij}^{h},$ and $\phi _{ij}^{0}$, which capture all the nontrivial signs
inherent to the model (without them, the model would simply reduce to a pure
bosonic one without any phase frustration). They are defined by
\begin{equation}
A_{ij}^{s}\equiv \frac{1}{2}\sum_{l\neq i,j}\left[ \theta _{i}(l)-\theta
_{j}(l)\right] \left( \sum_{\sigma }\sigma n_{l\sigma }^{b}\right) ~,
\label{as}
\end{equation}%
\begin{equation}
A_{ij}^{h}\equiv \frac{1}{2}\sum_{l\neq i,j}\left[ \theta _{i}(l)-\theta
_{j}(l)\right] n_{l}^{h}~,  \label{ah}
\end{equation}%
and%
\begin{equation}
\phi _{ij}^{0}\equiv \frac{1}{2}\sum_{l\neq i,j}\left[ \theta _{i}(l)-\theta
_{j}(l)\right] ~.  \label{a0}
\end{equation}%
The flux strengths of $A_{ij}^{s}$ and $A_{ij}^{h}$ are given in Eqs. (\ref%
{cond1}) and (\ref{cond2}), respectively, which are invariant under the
gauge transformations
\begin{equation}
h_{i}\rightarrow h_{i}e^{i\varphi _{i}}~,\text{ \qquad \quad }%
A_{ij}^{s}\rightarrow A_{ij}^{s}+(\varphi _{i}-\varphi _{j})~,  \label{u(1)1}
\end{equation}%
and
\begin{equation}
b_{i\sigma }\rightarrow b_{i\sigma }e^{i\sigma \theta _{i}}~,\text{ \qquad }%
A_{ij}^{h}\rightarrow A_{ij}^{h}+(\theta _{i}-\theta _{j})~.  \label{u(1)2}
\end{equation}%
Thus the t-J Hamiltonian in the phase string representation has an intrinsic
\textrm{U(1)}$\otimes $\textrm{U(1)} gauge structure and according to Eqs. (%
\ref{u(1)1}) and (\ref{u(1)2}), the holons and spinons will carry the gauge
charges of the gauge fields, $A_{ij}^{s}$ and $A_{ij}^{h}$, respectively. On
the other hand, the strengths of $A_{ij}^{s}$ and $A_{ij}^{h}$ are generated
by the local densities of the spinons and holons, respectively, according to
Eqs. (\ref{cond1}) and (\ref{cond2}). Finally $\phi _{ij}^{0}$ simply
describes a uniform $\pi $ flux per plaquette on a square lattice without
any dynamics.

Therefore, the \emph{full} sign structure hidden in the t-J model has been
explicitly sorted out and captured by the link variables, $A_{ij}^{s}$ and $%
A_{ij}^{h}$, as well as $\phi _{ij}^{0}$, in Eqs. (\ref{ht}) and (\ref{hj})
of the phase string/bosonization representation. At half-filling, $%
A_{ij}^{h} $ vanishes such that there is no nontrivial sign left in $H_{J},$
while $H_{t}=0$ under the constraint (\ref{single}). So the fermionic signs
of the electrons totally disappear here, and the residual intrinsic signs
only re-emerge as the holes are doped into the system, which are precisely
represented by the aforementioned topological gauge fields, and in this
sense the t-J model becomes an \emph{intrinsic} gauge model in the phase
string representation.

\subsection{Electron fractionalization}

Hence the phase string representation of the t-J model constitutes a
suitable starting point, as it smoothly connects the doping problem with the
undoped antiferromagnet -- the latter can be well described by the bosonic
RVB state without involving any sign problem\cite{lda_88,WZM_05}.

However, as pointed out in Ref. \cite{WZM_05}, a bosonic spinon\  \emph{%
defined} in the phase string representation, satisfying the constraint (\ref%
{single}), is not strictly charge-neutral: i.e., it is involved not only in
the \textquotedblleft neutral\textquotedblright \ superexchange process
described by $H_{J}$ [Eq. (\ref{hj})], but also in a \textquotedblleft
backflow\textquotedblright \ process\ accompanying holon (charge) hopping in
the $H_{t}$ term [Eq. (\ref{ht})]. As a matter of fact, it is shown\cite%
{WZM_05} that the bosonic RVB pairing is incompatible with the backflow
accompanying the holon hopping (see Sec. III C below), and thus it is
important to further distinguish these two different types of processes for
spinons in this particular representation.

\subsubsection{Two-component spinon description}

In order to properly accommodate these two distinct correlations of spins, a
neutral spinon Hilbert space has been introduced\cite{WZM_05} previously,
which satisfies the constraint $\sum_{\sigma }n_{i\sigma }^{b}=1$ instead of
Eq. (\ref{single}), even in the doped case. Such a \textquotedblleft
ghost\textquotedblright \ neutral spin state, denoted by $|\mathrm{RVB}%
\rangle ,$ can form a direct product state with the holon state $|\Phi
_{h}\rangle $
\begin{equation}
|\mathrm{RVB}\rangle \otimes |\Phi _{\mathrm{h}}\rangle .  \label{direct}
\end{equation}

Then the excessive spinons coinciding with the holon sites in Eq. (\ref%
{direct}) should be removed as they violate the Mott constraint (\ref{single}%
). A physical state constrained by Eq. (\ref{single}) may be then realized
by
\begin{equation}
\left \vert \Psi \right \rangle =\hat{\Pi}_{\mathrm{B}}|\mathrm{RVB}\rangle
\otimes |\Phi _{\mathrm{h}}\rangle  \label{physical}
\end{equation}%
with $\hat{\Pi}_{\mathrm{B}}=\cdot \cdot \cdot $ $(b_{l_{s}\sigma
_{s}}n_{l_{s}}^{h})(b_{l_{s+1}\sigma _{s+1}}n_{l_{s+1}}^{h})$ $\cdot \cdot
\cdot $ annihilating the unphysical neutral spinons at the holon sites. Here
$\hat{\Pi}_{\mathrm{B}}$ may be regarded as introducing a \emph{new} type of
spinons, namely, the \emph{backflow} spinons\cite{WZM_05}, in the
\textquotedblleft vacuum\textquotedblright \ state (\ref{direct}). Below we
show how to mathematically accurately incorporate such two-component spinon
building blocks into the phase string representation.

Define $a^{\dagger }$ as the creation operator for a backflow spinon, via a
one-to-one mapping
\begin{equation}
\left( b_{l\sigma }n_{l}^{h}\right) |\mathrm{RVB}\rangle \otimes |\Phi _{%
\mathrm{h}}\rangle \mapsto \hat{P}_{\mathrm{B}}\left[ \left( a_{l\bar{\sigma}%
}^{\dagger }e^{i\Xi _{l\bar{\sigma}}}\right) |0\rangle _{a}\otimes |\mathrm{%
RVB}\rangle \otimes |\Phi _{\mathrm{h}}\rangle \right]  \label{a-1}
\end{equation}%
where $|0\rangle _{a}$ denotes the vacuum state of the $a$-spinons and the
projector $\hat{P}_{\mathrm{B}}$ enforces the constraint in Eq. (\ref{PB}),
namely, the occupation number $n_{l\bar{\sigma}}^{a}$ of an $a$-spinon will
be always equal to the occupation number of the $b$-spinon in $|\mathrm{RVB}%
\rangle $ with the \emph{opposite} spin index $\bar{\sigma}=-\sigma $ at
site $l$, if and only if there is a holon sitting at the same site $l$ in $%
|\Phi _{h}\rangle $ (schematically it is represented by a dashed arrow in
Fig. \ref{Fig.2}). Here the phase factor $e^{i\Xi _{l\bar{\sigma}}}$ is
introduced to simplify the formulation, whose physical meaning will become
clear later.

Corresponding to this exact mapping in the Hilbert space, the t-J
Hamiltonian in the phase string representation can be further transformed as
follows.

\subsubsection{The hopping term}

Rewrite the hopping term $H_{t}$ in Eq. (\ref{ht}) as
\begin{equation}
H_{t}=-t\sum_{\langle ij\rangle \sigma }h_{j}\left( b_{i\sigma
}n_{i}^{h}\right) e^{i\phi _{ji}^{0}+i\sigma A_{ji}^{h}}\left(
n_{j}^{h}b_{j\sigma }^{\dagger }\right) h_{i}^{\dagger }e^{iA_{ij}^{s}}+h.c.
\label{ht0}
\end{equation}%
Further note that a shift $A_{ij}^{s}\rightarrow A_{ij}^{s}-\delta
A_{ij}^{s} $ has to be made, when the $b$-spinon Hilbert space is changed to
a neutral spinon basis satisfying $\sum_{\sigma }n_{i\sigma }^{b}=1$ instead
of Eq. (\ref{single}). Here $\delta A_{ij}^{s}=\frac{1}{2}\sum
\nolimits_{l\neq ij}\left[ \theta _{i}(l)-\theta _{j}(l)\right] \left(
\sum_{\sigma }\sigma n_{l\sigma }^{b}\right) n_{l}^{h}$ is obtained based on
the definition (\ref{as}). Then under the mapping (\ref{a-1}) the hopping
term is correspondingly changed to%
\begin{equation*}
H_{t}\mapsto \hat{P}\tilde{H}_{t}
\end{equation*}%
with%
\begin{equation}
\tilde{H}_{t}\equiv -t\sum_{\langle ij\rangle \sigma }\left( h_{i}^{\dagger
}h_{j}e^{iA_{ij}^{s}}\right) \left( a_{i\sigma }^{\dagger }a_{j\sigma
}e^{-i\phi _{ij}^{0}}\right) +h.c.  \label{ht1}
\end{equation}%
with $\hat{P}\equiv \hat{P}_{\mathrm{B}}\hat{P}_{s}$ where $\hat{P}_{s}$
enforces the constraint in Eq. (\ref{constraint=1}).

In obtaining Eq. (\ref{ht1}), the extra phases arising in Eq. (\ref{ht0})
has been absorbed by making the following choice:
\begin{eqnarray}
e^{i\Xi _{l\bar{\sigma}}} &=&e^{i\sigma \sum_{l\neq i}\theta
_{i}(l)n_{l\sigma }^{b}n_{l}^{h}}  \notag \\
&=&e^{i\sigma \sum_{l\neq i}\theta _{i}(l)n_{l\bar{\sigma}}^{a}}  \label{a-2}
\end{eqnarray}%
with utilizing Eq. (\ref{PB}). It is easy to verify that the phase factor $%
e^{i\Xi _{l\bar{\sigma}}}$ will serve as a 2D Jordan-Wigner operator to
precisely make the $a$-spinons defined in Eq. (\ref{a-1}) behave as \emph{%
fermions}. Note that in the earlier approach\cite{WZM_05}, without
introducing $e^{i\Xi _{l\bar{\sigma}}}$ to absorb the extra phase in Eq. (%
\ref{ht0}), the backflow spinons are treated in a boson representation which
is not as compact as in Eq. (\ref{ht1}).

Based on the precise mapping (\ref{a-1}), one has
\begin{eqnarray}
|\Psi \rangle &\mapsto &\hat{P}_{\mathrm{B}}\left[ |\Phi _{a}\rangle \otimes
|\mathrm{RVB}\rangle \otimes |\Phi _{\mathrm{h}}\rangle \right]  \notag \\
&=&\hat{P}\left[ |\Phi _{a}\rangle \otimes |\Phi _{b}\rangle \otimes |\Phi _{%
\mathrm{h}}\rangle \right]  \notag \\
&\equiv &\hat{P}|\tilde{\Psi}\rangle ,  \label{direct-2}
\end{eqnarray}%
where $|\Phi _{a}\rangle $ denotes the pure $a$-spinon state and $|\mathrm{%
RVB}\rangle \equiv \hat{P}_{s}|\Phi _{b}\rangle $. Generally $|\tilde{\Psi}%
\rangle $ here should be understood as a state expanded in terms of the
\emph{direct product bases} of the $a$-spinon, $b$-spinon, and $h$-holon.
Hence in the hopping term%
\begin{equation}
H_{t}|\Psi \rangle \mapsto \hat{P}\tilde{H}_{t}|\tilde{\Psi}\rangle ,
\label{ht2}
\end{equation}%
$\tilde{H}_{t}$ directly acts on the fractionalized state $|\tilde{\Psi}%
\rangle $ defined in Eq. (\ref{direct-2}).

Corresponding to the mapping in the Hilbert space as given in Eq. (\ref%
{direct-2}), the electron annihilation operator defined in Eq. (\ref{mutual}%
) in the phase string representation may be reexpressed by
\begin{equation}
c_{i\sigma }|\Psi \rangle \mapsto \hat{P}\left[ h_{i}^{\dagger }a_{i\bar{%
\sigma}}^{\dagger }\left( -\sigma \right) ^{i}e^{i\hat{\Omega}_{i}}\right] |%
\tilde{\Psi}\rangle  \label{c1}
\end{equation}%
which is obtained with using Eq. (\ref{a-2}), resulting in the phase shift
field $\hat{\Omega}_{i}=\left( \Phi _{i}^{s}-\Phi _{i}^{0}\right) /2$
defined in Eq. (\ref{phif})$.$

Similarly the spin operators $\mathbf{S}_{i}=\mathbf{S}_{i}^{b}$ in the
phase string representation [cf. Eqs. (\ref{sbz}) and (\ref{sb+})] can be
rewritten under the constraint Eq. (\ref{single}) as $\mathbf{S}_{i}=\mathbf{%
S}_{i}^{b}(1-n_{i}^{h})=\mathbf{S}_{i}^{b}-n_{i}^{h}\mathbf{S}_{i}^{b}$.
Then under the mapping of Eq. (\ref{direct-2}), it can be shown that
\begin{equation}
n_{i}^{h}\mathbf{S}_{i}^{b}\left \vert \Psi \right \rangle \mapsto -\hat{P}%
\mathbf{S}_{i}^{a}\left \vert \tilde{\Psi}\right \rangle ,  \label{s-ba}
\end{equation}%
with using Eqs. (\ref{a-1}) and (\ref{a-2}), such that%
\begin{equation}
\mathbf{S}_{i}\left \vert \Psi \right \rangle \mapsto \hat{P}\left[ \mathbf{S%
}_{i}^{b}+\mathbf{S}_{i}^{a}\right] |\tilde{\Psi}\rangle ,\text{\ }
\label{s}
\end{equation}%
where the $a$-spinon spin operators $\mathbf{S}_{i}^{a}$ are defined by%
\begin{equation}
S_{i}^{az}\equiv \frac{1}{2}\sum_{\sigma }\sigma a_{i\sigma }^{\dagger
}a_{i\sigma }  \label{saz}
\end{equation}%
and%
\begin{equation}
S_{i}^{a+}\equiv (-1)^{i}a_{i\uparrow }^{\dagger }a_{i\downarrow }\text{, \
\  \  \  \  \ }S_{i}^{a-}\equiv (-1)^{i}a_{i\downarrow }^{\dagger }a_{i\uparrow
},  \label{sa+}
\end{equation}%
(we assume the anticommutating relation, e.g., $a_{i\uparrow }^{\dagger
}a_{i\downarrow }=-a_{i\downarrow }a_{i\uparrow }^{\dagger }$, between the $%
a $-spinons of opposite spins without loss of generality).

\subsubsection{ The superexchange term}

For the superexchange term in Eq. (\ref{hj}), by introducing the factor $%
(1-n_{i}^{h})(1-n_{j}^{h})$ to explicitly enforce the no double occupancy
constraint (\ref{single}) in the enlarged Hilbert space $|\tilde{\Psi}%
\rangle $, one finds the following mapping

\begin{eqnarray}
H_{J}|\Psi \rangle &\longmapsto &-\frac{J}{2}\sum_{\langle ij\rangle }\hat{P}%
\left[ (1-n_{i}^{h})(1-n_{j}^{h})\left( \hat{\Delta}_{ij}^{s}\right)
^{\dagger }\hat{\Delta}_{ij}^{s}\right] |\tilde{\Psi}\rangle  \notag \\
&\equiv &\hat{P}\tilde{H}_{J}|\tilde{\Psi}\rangle \text{ \ },  \label{hj0}
\end{eqnarray}%
where%
\begin{eqnarray}
\tilde{H}_{J} &=&-\frac{J}{2}\sum_{\langle ij\rangle }\left( \hat{\Delta}%
_{ij}^{s}\right) ^{\dagger }\hat{\Delta}_{ij}^{s}-\frac{J}{2}\sum_{\langle
ij\rangle }n_{i}^{h}n_{j}^{h}\left( \hat{\Delta}_{ij}^{s}\right) ^{\dagger }%
\hat{\Delta}_{ij}^{s}  \notag \\
&&+\frac{J}{2}\sum_{\langle ij\rangle }\left( n_{i}^{h}+n_{j}^{h}\right)
\left( \hat{\Delta}_{ij}^{s}\right) ^{\dagger }\hat{\Delta}_{ij}^{s}\text{ \
.}  \label{hj2}
\end{eqnarray}%
$\tilde{H}_{J}$ in Eq. (\ref{hj2}) can be further reexpressed as%
\begin{eqnarray}
\tilde{H}_{J} &=&-\frac{J}{2}\sum_{\langle ij\rangle }\left( \hat{\Delta}%
_{ij}^{s}\right) ^{\dagger }\hat{\Delta}_{ij}^{s}-\frac{J}{2}\sum_{\langle
ij\rangle }\left( \hat{\Delta}_{ij}^{a}\right) ^{\dagger }\hat{\Delta}%
_{ij}^{a}  \notag \\
&&+J\sum_{\langle ij\rangle }\left( \mathbf{S}_{i}^{a}\cdot \mathbf{S}%
_{j}^{b}+\mathbf{S}_{i}^{b}\cdot \mathbf{S}_{j}^{a}\right) +JN_{h}
\label{hj1}
\end{eqnarray}%
in which in obtaining the second term on the right-hand-side (rhs) of the
first line, the relation
\begin{equation}
n_{i}^{h}n_{j}^{h}\left( \hat{\Delta}_{ij}^{s}\right) ^{\dagger }\hat{\Delta}%
_{ij}^{s}\mapsto \left( \hat{\Delta}_{ij}^{a}\right) ^{\dagger }\hat{\Delta}%
_{ij}^{a}  \label{delta-sa}
\end{equation}%
according to Eq. (\ref{a-1}) is used, and in obtaining the second line the
equality
\begin{equation}
\frac{1}{2}\left( \hat{\Delta}_{ij}^{s}\right) ^{\dagger }\hat{\Delta}%
_{ij}^{s}=-\mathbf{S}_{i}^{b}\cdot \mathbf{S}_{j}^{b}+1/4  \label{map2}
\end{equation}%
as well as Eq. (\ref{s-ba}) are utilized.

Therefore, by introducing a new kind of \textquotedblleft
backflow\textquotedblright \ $a$-spinon, the original $b$-spinon in the
phase string formalism will now always describe a \textquotedblleft
half-filled\textquotedblright \ neutral spin state constrained by Eq. (\ref%
{constraint=1}). Consequently two distinct processes, involving spins in the
hopping and superexchange terms of Eqs. (\ref{ht}) and (\ref{hj}), can be
mathematically depicted separately in terms of two kinds of spinons, as in
Eqs. (\ref{ht1}) and (\ref{hj1}). In this new formalism, the mutual
Chern-Simons gauge fields, $A_{ij}^{s}$ and $A_{ij}^{h}$, are still defined
by Eqs. (\ref{as}) and (\ref{ah}), but $A_{ij}^{s}$ is now always acting on
a half-filling background. The phase difference in $A_{ij}^{s}$ is then
absorbed into the backflow spinon, and naturally turns the latter into a
fermion. In the following, we will see that this kind of fractionalization
description will be very important for properly constructing a saddle-point
state at low doping.

\subsection{Mean-field scheme}

The t-J model has been first reformulated in the phase string representation
in order to accurately keep track of its peculiar and unique sign structure
and then expressed in a specific fractionalization formalism in terms of a
neutral bosonic spinon, a backflow fermionic spinon, and a bosonic holon, in
order to properly distinguish the microscopic hopping and superexchange
processes in the restricted Hilbert space. Now one is ready to construct an
effective theory/ground state based on the above new formulation.

The Schr\"{o}dinger equation $H_{t-J}|\Psi \rangle =E|\Psi \rangle $ can be
rewritten as
\begin{equation}
\hat{P}\left( \tilde{H}_{t-J}-E\right) |\tilde{\Psi}\rangle =0  \label{eg1}
\end{equation}%
by using $H_{t-J}|\Psi \rangle \longmapsto \hat{P}\tilde{H}_{t-J}|\tilde{\Psi%
}\rangle $ and $|\Psi \rangle \mapsto \hat{P}|\tilde{\Psi}\rangle $, where
\begin{equation}
\tilde{H}_{t-J}\equiv \tilde{H}_{t}+\tilde{H}_{J}  \label{htjtilde}
\end{equation}%
as defined in Eqs. (\ref{ht1}) and (\ref{hj1}).

Based on Eq. (\ref{eg1}), one may further make the following ansatz that
\begin{equation}
\tilde{H}_{t-J}|\tilde{\Psi}\rangle =E|\tilde{\Psi}\rangle  \label{eg2}
\end{equation}%
holds for both the ground state and low-lying excitation states. Generally
speaking, it is a sufficient but not a necessary condition for $\hat{P}|%
\tilde{\Psi}\rangle $ determined by Eq. (\ref{eg2}) to be an eigenstate of $%
\hat{P}\tilde{H}_{t-J}$. Later in Sec. III D we shall show that the
Bogoliubov quasiparticle excitation is indeed an exception, which satisfies
Eq. (\ref{eg1}) but not Eq. (\ref{eg2}), emerging as a collective mode
beyond the latter. But in the following we first focus on the solution of
Eq. (\ref{eg2}) and develop a generalized mean-field scheme.

Such a mean-field theory will be underpinned by a gauge-invariant bosonic
RVB order parameter%
\begin{equation}
\Delta _{ij}^{s}\equiv \left \langle \hat{\Delta}_{ij}^{s}\right \rangle
\neq 0,  \label{ds}
\end{equation}%
which was first introduced in Refs. \cite{weng_99,weng_07}. Then $\tilde{H}%
_{J}$ in Eq. (\ref{hj1}) can be linearized in terms of the \textquotedblleft
mean-field\textquotedblright \ order parameter $\Delta _{ij}^{s}$, giving
rise to an effective Hamiltonian $H_{s}$ in Eq. (\ref{hs}) with the order
parameter $\left( \Delta _{ij}^{s}\right) _{\mathrm{NN}}$ taken as
s-wave-like: $\left( \Delta _{ij}^{s}\right) _{\mathrm{NN}}=\Delta ^{s}$ and
$J_{s}=J\Delta ^{s}/2.$ Self-consistently one always finds the mean-field
\begin{equation}
\left \langle b_{j\sigma }^{\dagger }b_{i\sigma }e^{i\sigma
A_{ji}^{h}}\right \rangle _{\mathrm{NN}}=0  \label{bhop}
\end{equation}%
such that $\left( \Delta _{ij}^{s}\right) _{\mathrm{NN}}$ is the unique
order parameter for the bosonic RVB state\cite{weng_99,weng_07}. Equation (%
\ref{bhop}) implies that the bosonic RVB pairing is indeed incompatible with
the hopping, in contrast to the fermionic RVB case, which has been the basis
for introducing the backflow spinon to facilitate the hopping process in the
first place\cite{WZM_05}. It is noted that $H_{s}$ remains invariant, when $%
\left( \Delta _{ij}^{s}\right) _{\mathrm{NN}}$ is changed from s-wave to
d-wave-like, via a simple gauge transformation: $b_{i\sigma }\rightarrow $ $%
(-1)^{i_{y}}b_{i\sigma }$ so long as Eq. (\ref{bhop}) holds.

The kinetic energy of the $a$-spinons will arise from the hopping term $%
\tilde{H}_{t}$ in Eq. (\ref{ht1}), where a natural gauge-invariant
decoupling gives rise to a pure holon term $H_{h}$ in Eq. (\ref{hh}) and the
$a$-spinon hopping term in $H_{a}$ [i.e., the first term on the rhs of Eq. (%
\ref{ha})]. The rest of terms in $H_{a}$, including the pairing term for the
$a$-spinons, all come from $\tilde{H}_{J}$ in Eq. (\ref{hj1})$.$ The pairing
term in Eq. (\ref{ha}) is obtained by a conventional mean-field decoupling:%
\begin{equation}
\left( \hat{\Delta}_{ij}^{a}\right) ^{\dagger }\hat{\Delta}%
_{ij}^{a}\rightarrow \sum_{\sigma }\left( \Delta _{ij}^{a}\right) ^{\ast
}\sigma a_{i\sigma }^{\dagger }a_{j-\sigma }^{\dagger }+h.c.+...
\label{a-pair}
\end{equation}%
[a linearized $e^{-i\phi _{ij}^{0}}a_{i\sigma }^{\dagger }a_{j\sigma }$ term
is incorporated into the first term in Eq. (\ref{ha})], with a d-wave order
parameter $\Delta _{ij}^{a}\equiv \left \langle \Phi _{a}\right \vert \hat{%
\Delta}_{ij}^{a}|\Phi _{a}\rangle =\eta _{ij}\left \vert \Delta
_{ij}^{a}\right \vert $. Therefore, a \textquotedblleft
pseudogap\textquotedblright \ state characterized by the bosonic RVB pairing
of Eq. (\ref{ds}), which persists from half-filling to a finite doping, will
be described by an effective Hamiltonian $H_{\mathrm{eff}}=H_{s}+H_{h}+H_{a}$
as given in Eq. (\ref{heff}) [Sec. II B2].

It is important to point out that $H_{\mathrm{eff}}$ obtained based on Eq. (%
\ref{eg2}) is not strictly a conventional mean-field Hamiltonian. For
instance, at the mean-field approximation, one would have $%
t_{h}=t\sum_{\sigma }\left \langle a_{i\sigma }^{\dagger }a_{j\sigma
}e^{-i\phi _{ij}^{0}}\right \rangle \equiv tK^{a}$ and $t_{a}=t\left \langle
h_{i}^{\dagger }h_{j}e^{iA_{ij}^{s}}\right \rangle \equiv tH_{0}$, but under
the projection $\hat{P}$, one will have $t_{h}\sim t_{a}\sim t$ at low doping%
$.$ In principle, these parameters together with $J_{s}$ and $J_{a},$
appearing in Eqs. (\ref{hh}), (\ref{hs}), and (\ref{ha}), are not determined
by a standard self-consistent mean-field procedure. Instead, they will be
considered as the variational parameters, to be determined by minimizing the
ground state energy of $\hat{P}|\tilde{\Psi}_{\mathrm{G}}\rangle $ in Eq. (%
\ref{gsansatz}) with regard to the exact $H_{t-J}.$ Such a variational
scheme will decide the magnitudes and doping-dependences of these
parameters, but their detailed values are not crucial to the qualitative
physical consequences discussed in the present work.

\subsection{Quasiparticle as a collective excitation}

The above mean-field\ treatment of $\tilde{H}_{t-J}$ leads to an effective
Hamiltonian $H_{\mathrm{eff}}$ in Eq. (\ref{heff}), which determines the
fractionalized superconducting ground state (\ref{gsansatz1}). Based on $H_{%
\mathrm{eff}}$, two types of unconventional elementary excitations have been
identified, i.e., the spin-rotons and the fermionic $a$-spinons, which have
been discussed in Sec. II C1 and C2, respectively.

A quasiparticle mode, which carries both charge and spin-$1/2$ quantum
numbers, can be created by the electron operator defined in Eq. (\ref{c1}).
In the superconducting phase, with the holon condensation $\left \langle
h_{i}^{\dagger }\right \rangle \neq 0$ and the phase coherence $%
\left
\langle e^{i\hat{\Omega}_{i}}\right \rangle \neq 0$, an $a$-spinon
excitation may directly appear in the single-particle spectral function
according to Eq. (\ref{c1}). But such a decomposition structure is only a
part of the low-energy feature around the \emph{antinodal} region as
discussed in Sec. II C2 and C3.

In the following, we demonstrate that a more conventional Bogoliubov
quasiparticle will also emerge as an independent low-lying excitation in the
superconducting state, in addition to the above non-BCS-type excitations.
Being coherent around the nodal region, such a quasiparticle can be regarded
as a \emph{bound state} forming from the elementary building blocks of a
holon, an $a$-spinon, and a nonlocal phase shift based on Eq. (\ref{c1}).
Such a \textquotedblleft collective\textquotedblright \ mode goes beyond the
description of $H_{\mathrm{eff}}$ in Eq. (\ref{heff}), and will be correctly
described based on the \emph{original} full t-J Hamiltonian. The hopping
term in the latter will provide the necessary binding force for a stable
nodal quasiparticle, which has been previously shown\cite{WST_00,ZMW_03} in
the phase string representation by using the equation-of-motion method.
Below we give a similar proof based on the present decomposition given in
Eq. (\ref{c1}).

A quasiparticle excitation state may be generally constructed by%
\begin{equation}
\left \vert \mathbf{k}\sigma \right \rangle _{\mathrm{qp}}\equiv \alpha _{%
\mathbf{k}\sigma }^{\dagger }\left \vert \Psi _{\mathrm{G}}\right \rangle
\end{equation}%
where the creation operator $\alpha _{\mathbf{k}\sigma }^{\dagger }$ is a
linear combination of $c_{-\mathbf{k-}\sigma }$ and $c_{\mathbf{k}\sigma
}^{\dagger }$. The quasiparticle spectrum is given by%
\begin{eqnarray}
E_{\mathbf{k}} &=&\langle \mathbf{k}\sigma |H_{t-J}\left \vert \mathbf{k}%
\sigma \right \rangle _{\mathrm{qp}}-E_{\mathrm{G}}  \notag \\
&=&\langle \Psi _{\mathrm{G}}|\alpha _{\mathbf{k}\sigma }\left[ H_{\mathrm{%
t-J}},\alpha _{\mathbf{k}\sigma }^{\dagger }\right] |\Psi _{\mathrm{G}%
}\rangle
\end{eqnarray}%
using $\alpha _{\mathbf{k}\sigma }|\Psi _{\mathrm{G}}\rangle =0$ (assuming $%
|\Psi _{\mathrm{G}}\rangle $ being normalized).

For the t-J model, one generally has\cite{WST_00}
\begin{equation}
\left[ H_{t},c_{i\sigma }\right] =\frac{t}{2}(1+n_{i}^{h})\sum_{j=NN(i)}c_{j%
\sigma }+t\sum_{j=NN(i)}\left( c_{j\sigma }\sigma S_{i}^{z}+c_{j-\sigma
}S_{i}^{-\sigma }\right)  \label{eqomotion-t}
\end{equation}%
and
\begin{equation}
\left[ H_{J},c_{i\sigma }\right] =\frac{J}{4}c_{i\sigma
}\sum_{j=NN(i)}(1-n_{j}^{h})-\frac{J}{2}\sum_{j=NN(i)}\left( c_{i\sigma
}\sigma S_{j}^{z}+c_{i-\sigma }S_{j}^{-\sigma }\right) .  \label{eqomotion-j}
\end{equation}%
By acting them on the ground state in Eq. (\ref{gsansatz}) and using the
d-wave order parameters: $\Delta _{ij}^{SC}=\left \langle c_{i\uparrow
}c_{j\downarrow }\right \rangle $ and $K=\sum_{\sigma }\langle c_{i\sigma
}^{\dagger }c_{j\sigma }\rangle $ to linearize the rhs of the equations, one
finds%
\begin{equation}
\left[ H_{\mathrm{t-J}},c_{i\sigma }\right] |\Psi _{\mathrm{G}}\rangle
\simeq \left( t_{\mathrm{eff}}\sum_{j=NN(i)}c_{j\sigma }+\mu c_{i\sigma
}\right) |\Psi _{\mathrm{G}}\rangle -J\sum_{j=NN(i)}\Delta _{ij}^{\mathrm{SC}%
}\sigma c{_{j-\sigma }^{\dagger }}|\Psi _{\mathrm{G}}\rangle +\mathrm{decay}%
\text{ }\mathrm{term}\text{ }+\mathrm{scattering}\text{ \textrm{term}}
\label{eqomotion}
\end{equation}%
where $t_{\mathrm{eff}}=t(1+\delta )/2$ $+JK/4$ and $\mu $ is the chemical
potential.

If one can neglect the higher order terms on the rhs of Eq. (\ref{eqomotion}%
), including a \textquotedblleft decay term\textquotedblright \ and a
\textquotedblleft scattering term\textquotedblright \ to be given below,
then Eq. (\ref{eqomotion}) reduces to a linear equation in the $c$%
-operators, which can be diagonalized via the following Bogoliubov
transformation:
\begin{equation}
\alpha _{\mathbf{k}\sigma }^{\dagger }\propto u_{\mathbf{k}}c_{\mathbf{k}%
\sigma }^{\dagger }+\sigma v_{\mathbf{k}}c_{-\mathbf{k-}\sigma }
\label{bogo}
\end{equation}%
where $u_{\mathbf{k}}^{2}+v_{\mathbf{k}}^{2}=1$, with $u_{\mathbf{k}%
}^{2}=1+(\epsilon _{\mathbf{k}}-\mu )/E_{\mathbf{k}}$, $v_{\mathbf{k}%
}^{2}=1-(\epsilon _{\mathbf{k}}-\mu )/E_{\mathbf{k}}$, and $2u_{\mathbf{k}%
}v_{\mathbf{k}}=\Delta _{\mathbf{k}}/E_{\mathbf{k}}$. It leads to
\begin{equation}
\left[ H_{\mathrm{t-J}},\alpha _{\mathbf{k}\sigma }^{\dagger }\right] |\Psi
_{\mathrm{G}}\rangle =E_{\mathbf{k}}\alpha _{\mathbf{k}\sigma }^{\dagger
}|\Psi _{\mathrm{G}}\rangle  \label{egen}
\end{equation}%
with a BCS like spectrum%
\begin{equation}
E_{\mathbf{k}}=\sqrt{(\epsilon _{\mathbf{k}}-\mu )^{2}+\left( \Delta _{%
\mathbf{k}}\right) ^{2}}.  \label{Ek}
\end{equation}%
Here $\epsilon _{\mathbf{k}}=-2t_{\mathrm{eff}}(\cos k_{x}+\cos k_{y})$ and
\begin{equation}
\Delta _{\mathbf{k}}\equiv 2J\sum \limits_{\mathbf{q}}(\cos q_{x}+\cos
q_{y})\Delta _{\mathbf{k+q}}^{SC}  \label{G}
\end{equation}%
which can be easily shown to be d-wave if $\Delta _{\mathbf{k}}^{SC}$ is a
d-wave superconducting order parameter\cite{WST_00}. [Note that $\Delta _{%
\mathbf{k}}$ is scaled by an additional factor $\left \vert h_{0}\right
\vert ^{2}\propto \delta $ in Refs. \cite{WST_00,ZMW_03}, which is actually
an artifact in linearizing\cite{WST_00} the equation-of-motion.]

By noting that the chemical potential $\mu $ is given by requiring $%
\left
\langle \sum_{\mathbf{k\sigma }}c{_{\mathbf{k}\sigma }^{\dagger }}c_{%
\mathbf{k}\sigma }\right \rangle =(1-\delta )N$, one finds the quasiparticle
state $\alpha _{\mathbf{k}\sigma }^{\dagger }|\Psi _{\mathrm{G}}\rangle $
behaves similar to a Bogoliubov quasiparticle in a d-wave BCS state, where
it is built on a normal state with a large Fermi surface satisfying the
Luttinger theorem as decided by $\epsilon _{\mathbf{k}}=\mu .$ As a matter
of fact, in the above equation-of-motion calculation, if, say, the
next-nearest-neigbor hopping $t^{\prime }$ is added to $H_{\mathrm{t-J}}$, a
corresponding change in the band structure of $\epsilon _{\mathbf{k}}$ will
take place to give rise to a modified Fermi surface just like in an ordinary
band theory. By contrast, $t^{\prime }$ can be effectively renomalized to
zero in $H_{h}$ and $H_{a}$ due to the phase string effect, as to be shown
elsewhere, such that it will not directly affect the spin-roton and $a$%
-spinon excitation spectra. This further illustrates the distinction between
a fermionic $a$-spinon and a true Bogoliubov quasiparticle.

Hence a Bogoliubov quasiparticle is always a coherent excitation at a
sufficiently low energy as protected by the ODLRO (\ref{phcoh-0}). As
pointed out above, this Bogoliubov quasiparticle can be viewed as a bound
state in the fractionalized ground state, whose wave packet can eventually
break down at a higher energy in the superconducting state where the phase
coherence (\ref{phcoh-0}) is still maintained. To properly understand such a
stability, one needs to further inspect the higher order terms in Eq. (\ref%
{eqomotion}).

The \textquotedblleft decay term\textquotedblright \ in Eq. (\ref{eqomotion}%
) has the following leading contribution%
\begin{equation}
\mathrm{decay}\text{ }\mathrm{term}\text{ }\simeq \left \langle e^{i\hat{%
\Omega}}\right \rangle h_{0}^{\ast }\left[ Oa_{i\bar{\sigma}}^{\dagger
}(-\sigma )^{i}+\sum_{j=NN(i)}\left( P_{ij}a_{j\bar{\sigma}}^{\dagger
}-Q_{ij}a_{j\sigma }\right) (-\sigma )^{j}\right] |\Psi _{\mathrm{G}}\rangle
\label{decay term}
\end{equation}%
after a linearization with using the order parameters: $H_{0}$, $K^{a},$ $%
\Delta _{ij}^{a}$, and $h_{0}^{\ast }$, as well as $\left \langle e^{i\hat{%
\Omega}}\right \rangle $. Here the coefficients are given by $%
O=3tK^{a}-JH_{0},P_{ij}=\frac{1}{2}tH_{0}+\frac{3}{8}JK^{a}\sigma ,$ and $%
Q_{ij}=\frac{3}{4}J\Delta _{ij}^{a}.$ Such a decay term represents the
fractionalization tendency of a quasiparticle into an $a$-spinon, which is
already indicated by the decomposition form (\ref{c1}). It corresponds to a
coherent term so long as the phase coherence $\left \langle e^{i\hat{\Omega}%
_{i}}\right \rangle $ is maintained, in which the $a$-spinon appears at the
antinodal region, and thus in the momentum space is distinguished from the
Bogoliubov quasiparticle in the nodal region. Such a coherent term
disappears above $T_{c}$, even though the $a$-spinon can still remain
coherent at $T>T_{c}$.

Furthermore the scattering term reads
\begin{equation}
\mathrm{scattering}\text{ \textrm{term}}=\sum_{j=NN(i)}\left[ t\left(
c_{j\sigma }\sigma S_{i}^{bz}+c_{j-\sigma }S_{i}^{b-\sigma }\right) -\frac{J%
}{2}\left( c_{i\sigma }\sigma S_{j}^{bz}+c_{i-\sigma }S_{j}^{b-\sigma
}\right) \right] |\Psi _{\mathrm{G}}\rangle  \label{scatt}
\end{equation}%
which involves the scattering between the quasiparticle and the spin-roton
excitations composed of the $b$-spinons. By noting that a spin-roton
excitation has a finite \textquotedblleft resonance energy\textquotedblright
\ $E_{g}$ in the superconducting state (cf. Sec. II C1), the scattering in
Eq. (\ref{scatt}) may be safely neglected if one only focuses on the nodal
quasiparticles below this characteristic energy scale. On the other hand, at
an energy scale higher than $E_{g}$, a strong scattering between a
Bogoliubov quasiparticle and the background neutral spin excitations is
expected to dominate the single-particle spectral function in an appropriate
momentum region, which will be important to understand the ARPES data, but
is beyond the scope of the present paper.

\section{Conclusion and perspective}

The present mechanism for superconductivity resembles, by nature, what has
been proposed by Anderson\cite{pwa_87,pwa_11} that the spin RVB pairing is
turned into the Cooper pairing upon doping. But the basic structure of the
superconducting ground state, presented in Eq. (\ref{scgs-0}), is distinct
from the original proposal\cite{pwa_87} of the Gutzwiller-projected BCS
state in Eq. (\ref{BCS}) by that the neutral RVB and Cooper channels remain
clearly differentiated throughout the underdoped superconducting regime, as
illustrated by Figs. \ref{Fig.2} and \ref{Fig.1}, respectively. Such a
ground state has intrinsically embedded three essential types of
correlations: i.e., the AFM correlations in $|\mathrm{RVB}\rangle $; the
Cooper pairing of the doped holes; and a mutual influence/competition
between the two channels via a mutual statistical phase in $\Lambda _{h}$.

Consequently the superconductivity arises as a self-organization from these
correlations, rather than by default as in a Gutzwiller-projected BCS state
at finite doping. Without holes, the long-range AFM order will always win in
$|\mathrm{RVB}\rangle $, which provides a highly accurate description of the
ground state of the Heisenberg Hamiltonian. But a sufficient concentration
of doped holes will eventually turn the antiferromagnetism in $|\mathrm{RVB}%
\rangle $ into a true\ spin liquid state with short-range AFM correlations.
And by doing so the Cooper paired holes can gain phase coherence to realize
a high-$T_{c}$ superconductivity.

The key underlying such peculiar structure in the ground state can be
attributed to the singular phase shift introduced by the doped holes, which
is incorporated in Eq. (\ref{scgs-0}) by $\Lambda _{h}$. Such a
\textquotedblleft phase string\ effect\textquotedblright \ induced by the
hole hopping is irreparable, representing the exact sign structure of the
t-J model in the no double occupancy constrained Hilbert space, which
appears as mutual statistical signs drastically different from the
Fermi-Dirac statistical signs. Incorporating this new emergent statistics%
\cite{zaanen_09} is thus essential in order to correctly understand the non
Fermi liquid nature of doped Mott insulators.

Correspondingly there are three distinctive types of elementary excitations.
Firstly, the short-range RVB correlations in $|\mathrm{RVB}\rangle $ are
reflected by a \emph{finite} energy gap $E_{g}$ opened up in a neutral spin
excitation, called a spin-roton, which only reduces to the gapless spin wave
in an AFM state with a long-range RVB pairing, e.g., at half-filling. Such a
characteristic $E_{g}$ will protect the superconducting phase coherence from
phase disordering effect, caused by $\Lambda _{h}$ that closely monitors the
neutral spin correlations in $|\mathrm{RVB}\rangle $. In particular, it
determines\cite{MW_10} a unique $T_{c}$ formula (\ref{tc}). Secondly, the
fermionic excitations related to breaking up the Cooper pairs in Eq. (\ref%
{scgs-0}) are called backflow spinons ($a$-spinons) as they carry
well-defined spins and are therefore also contribute to spin excitations.
Such backflow spinon excitations will constitute a lower branch of spin
excitations, below $E_{g}$, as the size of the Cooper pairing is usually
larger than that of the RVB pairing. Thirdly, the $a$-spinon is distinct
from a conventional Bogoliubov particle by a singular\textrm{\ }phase shift
factor $e^{i\hat{\Omega}}$. Although the phase coherence $\left \langle e^{i%
\hat{\Omega}}\right \rangle \neq 0$ in the superconducting phase can make
the $a$-spinon show up in the single-particle channel around the antinodal
regime, the local singular fluctuations of $e^{i\hat{\Omega}}$ will in
addition induce a new collective mode as a bound state of the fractionalized
particles around the nodal regime. It is nothing but a conventional
Bogoliubov nodal quasiparticle.

Some additional unconventional properties can be inferred in such a d-wave
superconductor. In the spin channel, two branches of spin excitations near $%
(\pi $, $\pi ),$ separated by a resonancelike energy scale $E_{g},$ are
responsible by the spin rotons (upper branch) and $a$-spinons (lower
branch), respectively. In particular, strong magnetic fields should affect
the lower-branch first via the Zeeman effect, which can eventually destroy
the pairing of the $a$-spinons and thus the superconducting order parameter,
leading to small Fermi pockets formed by the fermionic $a$-spinons; On the
other hand, in the single-particle channel, a dichotomy of Bogoliubov
quasiparticle and the $a$-spinon coherent peaks can appear simultaneously in
the superconducting phase in the nodal and antinodal regions, respectively.
But the latter should disappear in the single-electron spectral function
once the phase of the superconducting order parameter is thermally
disordered by spin excitations in $|\mathrm{RVB}\rangle $, where each
excited spinon in $|\mathrm{RVB}\rangle $ automatically induces a current
vortex via $e^{i\hat{\Omega}}$, forming a spinon-vortex and proliferating in
a quantum vortex liquid state\cite{WQ_06}; Furthermore, in the presence of
weak magnetic fields, the Cooper pairing of doped holes provides a small
phase stiffness $\rho _{s}(0)\propto \delta $ at $T=0$. But $\rho _{s}$ will
be thermally reduced by the Bogoliubov nodal quasiparticles which couple to
the external \ electromagnetic fields with a full electric charge, leading
to $\rho _{s}(T)=$ $\rho _{s}(0)-aT$ with $a\sim O(1).$ Obviously all of
these will have strong experimental implications. But we shall further make
the detailed comparison with the cuprate superconductors elsewhere.

Finally, we remark that the present low-energy effective theory described by
$H_{\mathrm{eff}}$ in Eq. (\ref{heff}) resembles the so-called two-fluid
models, which have been phenomenologically proposed for the cuprate
superconductors\cite{pines_09,wen_05} and iron-based superconductors\cite%
{KLW_09} based on different theoretical considerations. The main similarity
with these approaches lies in that there exist both an itinerant BCS-type
component (i.e., $|\Phi _{a}\rangle $ in the present case) and a localized
spin liquid component (i.e., $|\mathrm{RVB}\rangle $ here). The main
distinction lies in how the local spin liquid component is mathematically
characterized: the full description of it in the present approach is
actually given by $|\mathrm{RVB}\rangle \otimes |\Phi _{h}\rangle $, which
is described by a mutual Chern-Simons gauge model [Eq. (\ref{psmodel})] that
respect spin rotation and time-reversal symmetries\cite{YTQW_11}. With the
doping effect self-consistently incorporated, such a local spin component
can naturally evolve from an AFM ordering state to a spin liquid state, and
exhibit a multilevel pseudogap behavior in the underdoped regime\cite%
{weng_07}.

\begin{acknowledgments}
I acknowledge useful discussions with P. W. Anderson, W.-Q. Chen, Z.-C. Gu,
S.-P. Kou, X.-L. Qi, D.-N. Sheng, C.-S. Tian, K. Wu, P. Ye, and J. Zaanen,
and in particular thank J.-W. Mei, V. N. Muthukumar, Y. Zhou for earlier
collaborations. This work was supported by NSFC No. 10834003, National
Program for Basic Research of MOST grant nos. 2009CB929402 and 2010CB923003.
\end{acknowledgments}

\appendix{}

\section{A hidden ODLRO in ground state (\ref{scgs})}

The ground state (\ref{scgs-0}) reduces to Eq. (\ref{scgs}) under the holon
condensation condition (\ref{hcond}). If the total number of holons is not
fixed, the ground state (\ref{scgs}) may be further expressed in a compact
form%
\begin{equation}
|\Psi _{G}\rangle =e^{\hat{D}}|\mathrm{RVB}\rangle .~
\end{equation}%
Since the operator $\hat{D}$ will introduce a pair of holes, satisfying%
\begin{equation}
\left[ \hat{N}_{h}\text{, }\hat{D}\right] =2\hat{D}  \label{D-1}
\end{equation}%
where $\hat{N}_{h}\equiv \sum_{i}n_{i}^{h}$, by using $\hat{N}_{h}\left(
\hat{D}\right) ^{n}|\mathrm{RVB}\rangle =2n\left( \hat{D}\right) ^{n}|%
\mathrm{RVB}\rangle $, one finds $\hat{N}_{h}|\Psi _{G}\rangle =2\hat{D}%
|\Psi _{G}\rangle $ such that
\begin{equation}
\left \langle \hat{D}\right \rangle \equiv \frac{\langle \Psi _{G}|\hat{D}%
|\Psi _{G}\rangle }{\langle \Psi _{G}|\Psi _{G}\rangle }=\frac{\langle \hat{N%
}_{h}\rangle }{2}\equiv \frac{\delta N}{2}  \label{D-3}
\end{equation}%
at a finite doping concentration $\delta $. The relation (\ref{D-3}) clearly
indicates that the ground state (\ref{scgs}) possesses a new ODLRO as given
in Eq. (\ref{D-0}) for a finite spatial separation of $ij.$


\begin{thebibliography}{99}
\bibitem{pwa_87} P. W. Anderson, Science \textbf{235}, 1196 (1987).

\bibitem{pwa_03} P. W. Anderson, P. A. Lee, M. Randeria, T. M. Rice, N.
Trivedi, and F. C. Zhang, J. Phys.: Condens. Matter \textbf{16}, R755
(2004), and references therein.

\bibitem{gros_07} For a review, see, B. Edegger, V.N. Muthukumar, and C.
Gros, Adv. Phys., \textbf{56}, 927 (2007).

\bibitem{LNW_06} For a review, see, P. A. Lee, N. Nagaosa, and X. G. Wen,
Rev. Mod. Phys. \textbf{78}, 17 (2006).

\bibitem{bza_87} G. Baskaran, Z. Zou, and P. W. Anderson, Solid State
Commun. \textbf{63}, 973 (1987).

\bibitem{zhang_88} F. C. Zhang, C. Gros, T. M. Rice, and H. Shiba,
Supercond. Sci. Technol. \textbf{1}, 36 (1988).

\bibitem{timusk_99} T. Timusk and B. Statt, Rep. Prog. Phys. \textbf{62}, 61
(1999).

\bibitem{shen_03} A. Damascelli, Z. Hussin, and Z.-X. Shen, Rev. Mod. Phys.
\textbf{75}, 473 (2003).

\bibitem{lda_88} S. Liang, B. Doucot, and P. W. Anderson, Phys. Rev. Lett.
\textbf{61}, 365 (1988).

\bibitem{weng_07} Z. Y. Weng, Intl. J. Mod. Phys. B \textbf{21}, 773 (2007);
arXiv:0704.2875.

\bibitem{zaanen_09} J. Zaanen and B. J. Overbosch, Phil. Trans. R. Soc. A
\textbf{369}, 1599 (2011); arXiv:0911.4070.

\bibitem{marshall_55} W. Marshall, Proc. Roy. Soc. (London) A\textbf{232},
48 (1955).

\bibitem{sheng_96} D. N. Sheng, Y. C. Chen, and Z. Y. Weng, Phys. Rev. Lett.
\textbf{77}, 5102 (1996).

\bibitem{weng_97} Z. Y. Weng, D. N. Sheng, Y.-Chen, and C. S. Ting, Phys.
Rev. B \textbf{55}, 3894 (1997).

\bibitem{WWZ_08} K. Wu, Z. Y. Weng, and J. Zaanen, Phys. Rev. B \textbf{77},
155102 (2008).

\bibitem{pwa-book} \emph{see}, P. W. Anderson, \emph{The Theory of
Superconductivity in the High $T_{c}$ Cuprates}, (Princeton Univ. Press,
Princeton, 1997).

\bibitem{WZM_05} Z.Y. Weng, Y. Zhou, V. N. Muthukumar, Phys. Rev. B \textbf{%
72}, 014503 (2005).

\bibitem{weng_99} Z. Y. Weng, D. N. Sheng, and C. S. Ting, Phys. Rev. B
\textbf{59}, 8943 (1999); Phys. Rev. Lett. \textbf{80}, 5401 (1998).

\bibitem{YTQW_11} P. Ye, C. S. Tian, X. L. Qi, and Z. Y. Weng, Phys. Rev.
Lett. \textbf{106}, 147002 (2011); S. P. Kou, X. L. Qi, and Z. Y. Weng,
Phys. Rev. B \textbf{71}, 235102 (2005).

\bibitem{sfermion} A. Auerbach and D. P. Arovas, Phys. Rev. Lett. \textbf{61}%
, 617 (1988); Sanjoy Sarker, C. Jayaprakash, H. R. Krishnamurthy, and
Michael Ma, Phys. Rev. B \textbf{40}, 5028 (1989); Daijiro Yoshioka, J.
Phys. Soc. Jpn. \textbf{58}, 32 (1989).

\bibitem{ZMW_03} Y. Zhou, V. N. Muthukumar, and Z. Y. Weng, Phys. Rev. B
\textbf{67}, 064512 (2003).

\bibitem{MW_02} V. N. Muthukumar and Z. Y. Weng, Phys. Rev. B \textbf{65},
174511 (2002).

\bibitem{WM_02} Z.Y. Weng and V.N. Muthukumar, Phys. Rev. B \textbf{66},
094509 (2002).

\bibitem{WQ_06} Z.Y. Weng and X.\ L. Qi, Phys. Rev. B \textbf{74}, 144518
(2006).

\bibitem{MW_10} J. W. Mei and Z. Y. Weng, Phys. Rev. B \textbf{81}, 014507
(2010).

\bibitem{Uemura} Y.J. Uemura, J. Phys. Condens. Matter \textbf{16}, S4515
(2004); Y.J Uemura, Physica B \textbf{374-375}, 1 (2006).

\bibitem{WST_00} Z. Y. Weng, D. N. Sheng, and C. S. Ting, Phys. Rev. B
\textbf{61}, 12328 (2000).

\bibitem{pwa_11} P. W. Anderson, arXiv:1011.2736.

\bibitem{pines_09} V. Barzykina and D. Pines, Adv. Phys. \textbf{58}, 1
(2009), and references therein.

\bibitem{wen_05} T. C. Ribeiro and X. G. Wen, Phys. Rev. B \textbf{74},
155113 (2006); Phys. Rev. Lett., \textbf{95, }057001 (2005).

\bibitem{KLW_09} S. P. Kou, T. Li, and Z. Y. Weng, EPL \textbf{88,} 17010
(2009).
\end{thebibliography}
\end{document}